\documentclass{emulateapj}

\usepackage{apjfonts}
\usepackage{lscape}
\usepackage{amsmath}

\newcommand{\ltappeq}{\raisebox{-0.6ex}{$\,\stackrel{\raisebox{-.2ex}{$\textstyle <$}}{\sim}\,$}}
\newcommand{\gtappeq}{\raisebox{-0.6ex}{$\,\stackrel{\raisebox{-.2ex}{$\textstyle >$}}{\sim}\,$}}

\shorttitle{Stellar Exotica in 47 Tuc}
\shortauthors{Knigge et al.}

\begin{document}

%% LaTeX will automatically break titles if they run longer than
%% one line. However,you may use \\ to force a line break if
%% you desire.

\title{Stellar Exotica in 47 Tucanae\footnotemark[1]}
\footnotetext[1]{Based on observations made with the NASA/ESA Hubble Space Telescope,
obtained at the Space Telescope Science Institute, which is operated
by the Association of Universities for Research in Astronomy, Inc., 
under NASA contract NAS 5-26555.}

%% Use \author, \affil, and the \and command to format
%% author and affiliation information.
%% Note that \email has replaced the old \authoremail command
%% from AASTeX v4.0. You can use \email to mark an email address
%% anywhere in the paper, not just in the front matter.
%% As in the title, use \\ to force line breaks.

%\author{Christian Knigge\altaffilmark{2}, Andrea Dieball\altaffilmark{2}}
%\altaffiltext{2}{School of Physics \& Astronomy, University of Southampton, Southampton SO16 7AR, UK}
%\author{Jes\'{u}s Ma\'{i}z Apell\'{a}niz\altaffilmark{3,4}}
%\altaffiltext{3}{Instituto de Astrof\'{\i}sica de Andaluc\'{\i}a-CSIC, 18008 Granada, Spain}
%\altaffiltext{4}{Ram\'on y Cajal Fellow}
%\author{Knox S. Long\altaffilmark{5}}
%\altaffiltext{5}{Space Telescope Science Institute, Baltimore, MD 21218, USA}
%\author{David R. Zurek\altaffilmark{6}}
%\author{Michael M. Shara\altaffilmark{6}}
%\altaffiltext{6}{American Museum of Natural History, New York, NY 10024, USA}

\author{Christian Knigge, Andrea Dieball}
\affil{School of Physics \& Astronomy, University of Southampton, Southampton SO16 7AR, UK}
\author{Jes\'{u}s Ma\'{i}z Apell\'{a}niz}
\affil{Ram\'on y Cajal Fellow}
\affil{Instituto de Astrof\'{\i}sica de Andaluc\'{\i}a-CSIC, 18008 Granada, Spain}
\author{Knox S. Long}
\affil{Space Telescope Science Institute, Baltimore, MD 21218, USA}
\author{David R. Zurek, Michael M. Shara}
\affil{American Museum of Natural History, New York, NY 10024, USA}

%% Mark off your abstract in the ``abstract'' environment. In the manuscript
%% style, abstract will output a Received/Accepted line after the
%% title and affiliation information. No date will appear since the author
%% does not have this information. The dates will be filled in by the
%% editorial office after submission.

\begin{abstract}

We present far-ultraviolet (FUV) spectroscopy obtained with the {\em
Hubble Space Telescope (HST)} for 48 blue objects in
the core of 47~Tuc. Based on their position in a FUV-optical
colour-magnitude diagram, these were expected to include cataclysmic
variables (CVs), blue stragglers (BSs), white dwarfs (WDs) and other
exotic objects.  For a subset of these sources, we also construct
broad-band, FUV through near-infrared spectral energy distributions. 
Based on our analysis of this extensive data set, we
report the following main results. (1) We detect emission lines in three
previously known or suspected CVs and thus spectroscopically confirm
the status of these systems. We also detect new dwarf nova eruptions
in two of these CVs. (2) Only one other source in our spectroscopic
sample exhibits marginal evidence for line emission. Thus CVs are not
the only class of objects found in the gap between the WD and main
sequences, nor are they common amongst objects near
the top of the WD cooling sequence. Nevertheless, predicted and
observed numbers of CV agree to within a factor of about
2-3. (3) We have discovered a hot
($T_{eff} \simeq 8700$~K), low-mass ($M \simeq 0.05~M_{\odot}$)\ 
secondary star in a previously known 0.8~d binary system. This
exotic object completely dominates the binary's FUV-NIR output and is
probably the remnant of a subgiant that has been stripped of its
envelope. Since this object must be in a short-lived evolutionary
state, it may represent the  ``smoking gun'' of a recent dynamical
encounter. (4) We have found a Helium WD, only the
second such object to be optically detected in 47~Tuc, and the first
outside a millisecond pulsar system. (5) We have discovered a bright
BS with a young WD companion, the only BS-WD binary known in any
GC. (6) We have found two additional candidate WD binary systems, one
containing a  MS companion, the other containing a subgiant. (7) We
estimate the WD binary fraction in the core of 47~Tuc to be  
$15\%~^{+17\%}_{-9\%}~{\rm (stat)}~^{+8\%}_{-7\%}~{\rm (sys)}$. (8) The 
mass of the optically brightest BS in our sample may exceed twice the
cluster turn-off mass, but the uncertainties are too large for this to be
conclusive. Thus there is still no definitive example of such a
``supermassive'' BS in any GC.  Taken as a whole, our study
illustrates the wide range of stellar exotica that are lurking in the
cores of GCs, most of which are likely to have undergone significant
dynamical encounters.

\end{abstract}

%% Keywords should appear after the \end{abstract} command. The uncommented
%% example has been keyed in ApJ style. See the instructions to authors
%% for the journal to which you are submitting your paper to determine
%% what keyword punctuation is appropriate.

\keywords{
globular clusters: individual(NGC 104) --- 
blue stragglers --- 
novae, cataclysmic variables --- 
white dwarfs -- 
binaries:close ---  
ultraviolet: stars --- 
techniques: spectroscopic
}

\section{Introduction}

Globular clusters (GCs) harbour an impressive array of exotic stellar 
populations, such as blue stragglers (BSs), millisecond pulsars
(MSPs), X-ray binaries (XRBs), cataclysmic variables (CVs) and Helium
white dwarfs (He WDs). The sizes of
these populations can be significantly enhanced in GCs, relative to
the Galactic field. This is a direct result of the high stellar
densities encountered in GC cores (up to at least $10^{6} \, M_{\odot}\,{\rm
  pc}^{-3}$). In such extreme environments, dynamical interactions
between cluster members occur rather frequently, which opens up new
and efficient production channels for virtually all types of ``stellar
exotica.'' Thus GCs are excellent laboratories for studying these
otherwise extremely rare objects (Maccarone \& Knigge 2007). 

This argument can also be turned around. Since exotic stellar
populations in GCs are preferentially produced by dynamical 
interactions, they can be used as tracers of a GC's dynamical
state. For example, the number of faint X-ray sources in GCs has been
shown to correlate well with the 
predicted stellar encounter rate (Pooley et al. 2003; Heinke et
al. 2003a; Gendre, Barret \& Webb 2003a; Pooley \& Hut 2006), and
deviations from this relationship can be related to specific 
dynamically-relevant factors (e.g. the highly elliptical orbit of
NGC~6397; see Pooley et al. 2003). 

However, exotic stellar populations are not merely passive tracers of
GC evolution. Instead, most of these populations are composed of close
binaries that actively drive the evolution of their host clusters
towards evaporation. Both primordial and dynamically-formed close
binaries can be thought 
of as heat sources in this context: in successive dynamical 
encounters with passing single stars, close binaries tend to give up
gravitational binding energy to their interaction partners, and hence
to their parent cluster as a whole. The binaries themselves become
increasingly tight in the process (i.e. they ``harden''). Thus stellar
exotica in GCs are not merely {\em formed} dynamically, but are part
of the central feedback loop that links the dynamical evolution of a GC
to the formation and evolution of its close binary population
(e.g. Hut et al. 1992 [Section 3.1.3]; Shara \& Hurley 2006; Hurley,
Aarseth \& Shara 2007).

\begin{figure*}
\epsscale{1.15}
\center
\includegraphics[scale=0.55,angle=90,keepaspectratio=true]{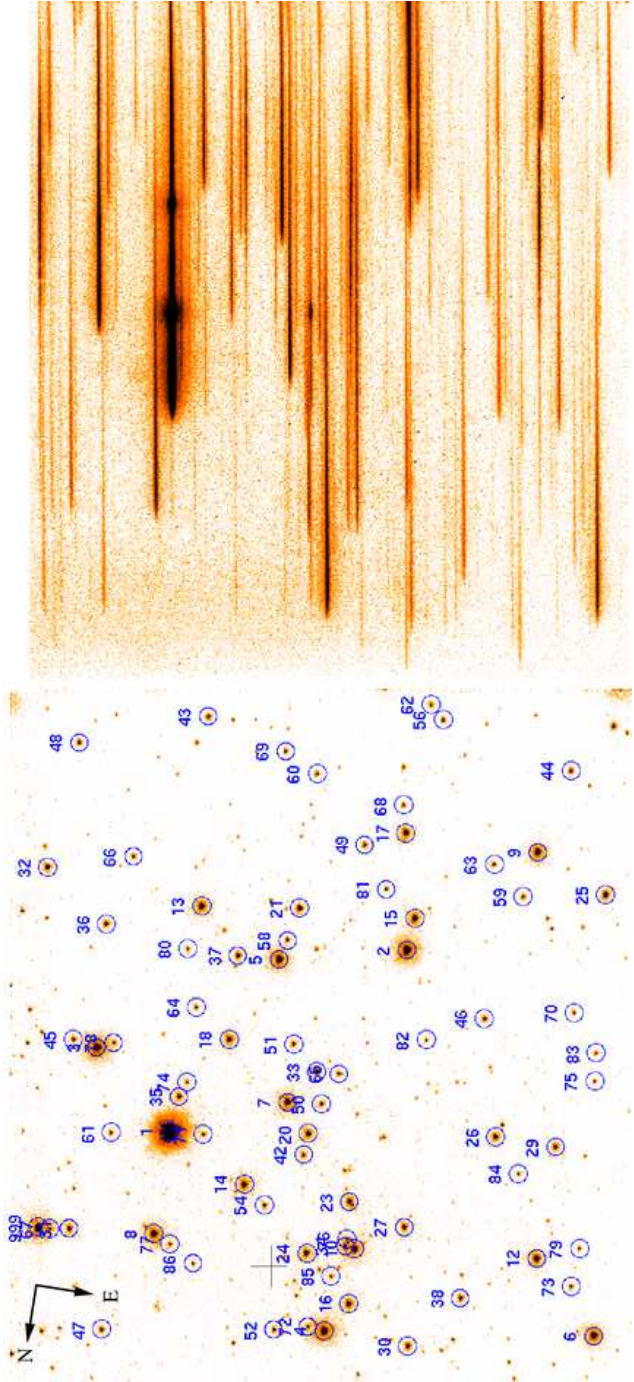}
%\plotone{f1.eps}
\caption{{\em Left Panel:} The direct FUV image of a roughly 
26\arcsec$\times$23\arcsec\ region in the core of 47~Tucanae. Sources 
bright enough to be included in the spectral extraction are circled
and labelled with a spectroscopic identification number. The
cross marks the approximate position of the cluster center
(Guhathakurta et al. 1992). Essentially all of this image is inside
the $R_{c} = 24$\arcsec\ core radius of the cluster (Howell,
Guhathakurta \& Gillilan 2000). {\em Right
  Panel:} The slitless FUV spectral image of 
the same region. Each trail in the image is the dispersed spectrum of
a bright FUV source. Note the obvious emission lines in the spectrum
of the brightest source (Star~1 = AKO~9; c.f. Knigge et al. 2003).}
\label{fig:images}
\end{figure*}

Even though exotic stellar populations are abundant in GCs compared to
the Galactic field, they still represent only a small fraction of the
total number of stars in any cluster. Thus even in GCs, finding and
characterizing these populations is a challenging task. One 
useful feature common to most members of these populations is that
their spectral energy distributions (SEDs) tend to be ``bluer'' than
those of ordinary stars (i.e. a 
larger fraction of their radiation is emitted at higher
frequencies). This can be exploited. For example, deep X-ray
surveys have turned out to be a powerful method for locating 
XRBs and CVs in GCs (e.g. Grindlay et al. 2001ab; Pooley et al. 2002;
Heinke et al. 2003b; Gendre, Barret \& Webb 2003b; Heinke et al. 2005;
Webb, Wheatley \& Barret 2006). Similarly, far-ultraviolet (FUV)
surveys of GCs are sensitive to an even wider range
of ``hot'' exotic populations, while still avoiding the crowding
problems associated with optical searches (e.g. Ferraro et al. 2001;
Knigge et al. 2002 [hereafter Paper I]; Dieball et al. 2005a; Dieball
et al. 2007). However, the detection of 
X-ray emission and/or FUV excess is usually not enough to yield a
unique source classification. Thus additional information -- ideally
based on spectroscopic observations -- is required for a full census
of the exotic stellar populations in GCs. 

Here, we present FUV spectroscopy obtained with the {\em Hubble Space
  Telescope (HST)} of 48 FUV-excess sources in the 
core of 47 Tuc. Based mainly on their position in a FUV/optical
colour-magnitude diagram (CMD), these sources are expected to include
BSs, CVs and young WDs (Paper I). We find examples of all of
these classes in our spectroscopy, along with unexpected discoveries
of even more exotic objects. In Section~2, we describe the 
observations we use in our study, along with our data reduction and  
analysis methods. We also highlight the positions of our spectroscopic
targets in the CMD and present proper motion-based membership
information for a subset of sources. In Section~3, we present an
overview of all the individual spectra and assess the overall
relationship between CMD position and spectroscopic
classification. The heart of this paper is 
Section~4, where we take a detailed look at the most interesting
sources. We also construct global FUV through near-infrared (NIR) 
SEDs for these objects and use joint fits to the spectroscopic
and SED data to determine their nature. In Section~5, we discuss some
of the wider implications of our work and estimate the WD binary
fraction in 47~Tuc. Finally, in Section~6, we summarize our main
results and conclusions. 

%8 BS
%8 gap
%9 WDs
%23 no-opt

\section{Observations and Data Analysis}

\subsection{Far-Ultraviolet Imaging and Slitless Spectroscopy}
\label{sec:fuvdata}

We have carried out a deep spectrosopic and photometric survey of 47
Tuc in the FUV waveband. In total, we obtained 30 orbits of
{\em HST/STIS} observations for our program (G0-8279), split into 6 epochs
of 5 orbits each. Imaging exposures were obtained at the beginning and
end of each epoch, with the rest of the time being spent on slitless
spectroscopy. Exposure times were typically 600~s for both imaging and
spectroscopy. In total, we obtained approximately 14,600~s of FUV
imaging and 82,200~s of slitl<ess spectroscopy. 

All of the observations used the FUV-MAMA detectors, together with the
F25QTZ filter. This filter blocks geocoronal Ly $\alpha$\, O{\sc i}
1304 \AA\, and O{\sc i}] 1356 \AA\ emission, which would otherwise
produce an unacceptably high background in our slitless
spectral images. The resulting photometric bandpass is strongly
asymmetric, with peak transmission at about 1487~\AA, just above the
sharp cut-off around 1450~\AA; the pivot wavelength is 1595~\AA, and
the full-width at half-maximum is 229~\AA. The 1024$\times$1024 pixel
FUV-MAMA detectors cover a field of view of about
25\arcsec$\times$25\arcsec\ (corresponding to about 1/3 of 47~Tuc's
core) with a spatial sampling of about 24.5~mas~pix$^{-1}$. 

The data reduction and analysis steps carried out for the direct
images -- which include the construction of a combined master image, as
well as source detection and photometry -- have already  been described
in Paper I. We will use the resulting FUV source catalogue throughout
this paper, except for three small changes and 
additions. First, in Paper I, we restricted our attention to the 
spatial area common to {\em all} direct images. Even though the
spatial shifts between images taken in different epochs are quite
small, this restriction removed one bright FUV source for which we now
have a high-quality spectrum. We have therefore added this source --
Star 999 in the notation adopted below -- to the photometric
catalogue. Second, we have applied a small (0.086~mag) correction to
the FUV magnitudes obtained in Paper I in order to account for the
sensititivity loss of the FUV-MAMA detectors at the time of the
observations. This correction was not available at the time of Paper I's
publication. Third, as described in Knigge et al. (2006), we have
applied distortion and boresight corrections to the FUV source
positions. The boresight correction is designed to put our equatorial 
coordinates into the absolute astrometric frame defined by the Chandra
source positions listed in Heinke et al. (2005). 

Our slitless FUV spectroscopy was carried out with the G140L grating,
which provides a dispersion of 0.584~\AA\ pix$^{-1}$ and a spectral
resolution of about 1.2~\AA. This set-up can cover a spectral
range of roughly 1450~\AA\ -- 1800~\AA. However, note that in slitless
observations, the 
actual wavelength range covered for each source depends on its spatial
position on the detector. Since the observatory software automatically
applied a spatial offset to all spectroscopic observations, the
effective field of view for the spectral images was only about 85\% of
the field available in the direct images. 

Figure~\ref{fig:images} shows the co-added direct and spectral images
side-by-side. This figure illustrates the high efficiency of the
slitless approach, but also the technical challenges it poses. By
moving to the FUV waveband, we can drastically 
reduce the crowding that plagues optical images, since the vast
majority of ``normal'' cluster 
members are simply not hot enough to produce appreciable amounts of
FUV flux (see Figure~1 in Paper I for a graphic illustration of this
effect). As a result, multi-object slitless spectroscopy of dense GC
cores becomes feasible in the FUV and provides an efficient technique
for determining the nature of exotic stellar
populations. However, extracting single-source spectra from slitless
spectral images like that shown in Figure~\ref{fig:images} is not a
trivial task. Part of the difficulty stems from the fact that the 
spatial and spectral dimensions are not independent. However, the more 
important problem is the blending of sources in the spectral image. 

We deal with these technical challenges by using the {\sc multispec}
software package, which has been specifically designed to facilitate
the extraction of single-source spectra from multi-object slitless
spectral images (Ma\'{i}z-Apell\'{a}niz 2005, 2007). Briefly, {\sc 
multispec} uses the 
position, brightness and colour of each source (as determined from
direct images of the field) to generate an initial estimate of its 
spectrum. Next, all of these estimates are combined to construct 
a complete synthetic spectral image that can be directly compared (in a
$\chi^2$ sense) to the actual spectral image. The initial estimates of
the individual spectra 
are then iteratively optimized until the best possible match between
synthetic and real spectral images has been achieved. 

A few points are worth noting regarding this process. First, a key 
factor in the construction of the synthetic spectral image from the
individual spectra is the wavelength-dependent cross-dispersion
profile (CDP). This is effectively the point spread function (PSF) 
perpendicular to the dispersion direction in the spectral image. We
derived the CDP for our instrumental set-up from
calibration observations in the {\em HST} archive. Second, blending is
self-consistently accounted for in the extraction, since an arbitrary
number of sources can contribute to any given pixel in the synthetic 
spectral image. Third, the extraction
process also allows for a smoothly varying global detector background,
which is iteratively estimated from the residuals between real and
synthetic spectral images. Fourth, we extract source
spectra separately for each epoch. This avoids any problems associated
with the slight spatial offsets between the spectral images for each
epoch. Our final spectrum for each source is then the
exposure-weighted average of the optimized single-epoch spectra. 

\begin{figure} 
%\epsscale{1.2}
\epsscale{1.0}
\plotone{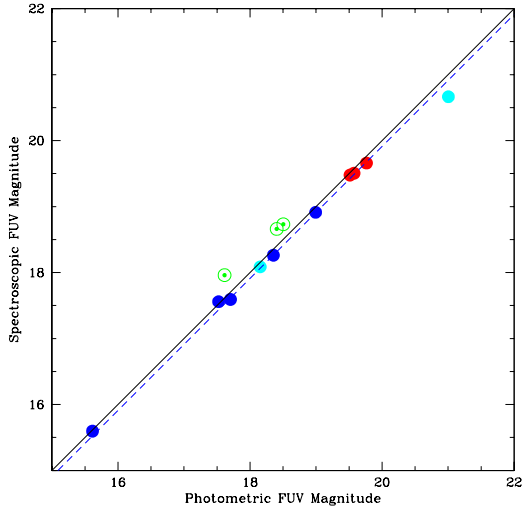}
\caption{A comparison of spectroscopically and photometrically
estimated FUV magnitudes for several bright FUV sources (see text for
details). The photometric magnitudes include the 0.086~mag correction
for the time-dependent sensitivity loss of the FUV-MAMA detectors (see
Section~\ref{sec:fuvdata}). Colours mark 
different source types, based on the photometric classification scheme
shown in Figure~\protect\ref{fig:cmd} and described in
Section~\protect\ref{sec:cmd}. Gap objects are shown in blue, white
dwarfs in red, blue stragglers in green, and sources without optical
counterparts in cyan. The solid black line corresponds to $x = y$, the
dashed black line corresponds to the best-fitting line of slope unity
to all data points except the blue stragglers. The $y$-intercept of
this line is -0.089 mag.}
\label{fig:fluxcal}
\end{figure}

The extracted source spectra will be good approximations to the true
source spectra, as long as (i) the CDP is sufficiently accurate, 
(ii) the input source catalogue is sufficiently accurate and complete,
and (iii) blending is not too severe. All of these conditions imply
that there is a limit to the ability of the algorithm in extracting
spectra for faint objects that are heavily blended with one or more
bright objects. In this situation, even very slight errors in the
extraction of the bright source spectra can lead to large systematic
errors in the extracted faint source spectra. 

In order to flag areas of a spectrum that might be compromised by 
severe blending, we calculate a wavelength-dependent ``blend ratio''
for each extracted source. This is defined as the ratio of counts from 
the source itself to the counts produced by all other extracted
objects, as estimated at the peak of the source CDP. We stress that
this blend ratio is no panacea. In particular, since our estimate of
it is necessarily 
based on the extracted source spectra, it can itself be affected by  
blending. For example, if the flux of a faint source has been
overestimated due to blending, the blend ratio associated with that
source will be underestimated. Nevertheless, the blend ratio does
provide an extremely useful, if qualitative, measure of the
wavelength-dependent reliability of each source spectrum. As a rule of
thumb, blending is not a serious issue whenever the blend ratio is
$\ltappeq 0.1$, whereas spectral regions with blend ratios $\gtappeq
1$\  may be severely compromised.

% THIS IS WHERE F3 GOES IN EMULATEAPJ

In order to test our spectral extraction method, we have folded the
extracted spectra of several bright FUV sources with particularly good
spectral coverage through the STIS/FUV-MAMA/F25QTZ photometric 
response curve. This was done with the {\sc synphot} package within
{\sc iraf/stsdas} and yielded spectroscopic estimates of the FUV
magnitudes for these sources. These could be directly compared to the
photometric magnitudes, as shown in Figure~\ref{fig:fluxcal}. The
agreement between spectroscopic and photometric estimates is generally
quite good, except for the 3 BSs in the sample. These generally have
spectroscopic magnitudes that are systematically fainter than the
photometric ones, by 0.2~mag~-~0.3~mag. This disagreement is actually
to be expected, since we can only extract spectra out to
1800~\AA. Beyond this, the throughput drops dramatically and is not well
calibrated. However, the FUV spectra of BSs rise extremely steeply
towards longer 
wavelengths (see Section~\ref{sec:bs_specs}), and the FUV-MAMA/F25QZ
photometric bandpass does retain some slight sensitivity there. Thus
wavelengths beyond our spectrosopic cut-off  
contribute non-negligibly for the BSs. For the other sources,
photometric and spectroscopic estimates agree quite well. The rms
scatter is only about 0.1~mag, which is acceptable, especially considering that our 
sample includes known variables (such as the CVs AKO~9, V1 and V2; see
Section~\ref{sec:gap_specs} below). There is marginal evidence
for a slight offset of 0.089 mag (in the sense of the spectroscopic 
estimates being brighter), so we apply this as a uniform correction to
all of our extracted spectra.

In total, 77 FUV sources were included in the spectral extraction
process and are highlighted in Figure~\ref{fig:images}. Useful spectra
could be obtained for 48 of these sources, and these are listed in
Table~\ref{tab:specdat} along with their basic properties.

\subsection{Optical Imaging}
\label{sec:optdata}

The FUV-optical CMD presented in Paper~I served as a starting point in 
the classification of our spectroscopic targets (see
Section~\ref{sec:overview}). The optical photometry for this CMD was
based on a deep, co-added WFPC2/PC/F336W (roughly U-band) image of the
cluster core and has already been described in Paper~I. 

For the subset of FUV sources discussed in Section~\ref{sec:sed}, 
we additionally constructed broad-band, FUV through NIR SEDs. With the
exception of the FUV data point (which comes 
from our own STIS/F25QTZ direct image), all of this photometry was
obtained from ACS/HRC observations of 47 Tuc in the {\em HST} archive. These
observations spanned the full complement of ACS/HRC broad-band
filters, including F250W ($\lambda_p = $\ 2716~\AA), F330W ($\lambda_p
=\ $3363~\AA), F435W ($\lambda_p = $\ 4311~\AA), F475W ($\lambda_p = 
$\ 4776~\AA), F555W ($\lambda_p = $\ 5256~\AA), F606W ($\lambda_p =
$\ 5888~\AA), F625W ($\lambda_p = $\ 6296~\AA), F775W ($\lambda_p = 
$\ 7665~\AA), F814W ($\lambda_p = $\ 8115~\AA), and F850LP ($\lambda_p = 
$\ 9145~\AA); here, $\lambda_p$\ is the pivot wavelength of each
filter. Due to the limited field of view and depth of the exposures we
used, only a subset of these filters are usually available for each
target.

All of our ACS/HRC photometry was carried out with Andrew Dolphin's
PSF-fitting package {\sc dolphot}
\footnote{http://purcell.as.arizona.edu/dolphot}, 
which is a
modified version of the WFPC2-optimized {\sc HSTphot} 
package (Dolphin 2000). The advantage of {\sc dolphot} for our
purposes is that it provides an ACS-optimized mode, in which
detector- and filter-specific simulated {\sc Tiny Tim} PSFs are used as
baseline PSF
models\footnote{http://www.stsci.edu/software/tinytim/tinytim.html}.  
In ACS mode, {\sc dolphot} applies CTE corrections automatically and
provides fully calibrated photometric measurements.

Throughout this paper, all magnitudes are given on the STMAG system,
where 
\begin{equation}
m_{STMAG} = -2.5 \log{F_{\lambda}} - 21.1. 
\label{eq:stmag}
\end{equation}
For photometric measurements, $F_{\lambda}$\ is the (constant) flux 
of a flat-spectrum source that would produce the observed count rate
in the photometric band-pass. In Section~4, we will also adopt the
STMAG system for spectroscopic data, by  using
Equation~\ref{eq:stmag}\ as a monochromatic definition of
$m_{STMAG}$. This convention allows us to put spectroscopic and 
photometric measurements onto a common, easily interpretable scale.

\subsection{The FUV-optical Colour-Magnitude Diagram}
\label{sec:cmd}

\begin{figure*}
%\epsscale{1.1}
\epsscale{1.0}
\plotone{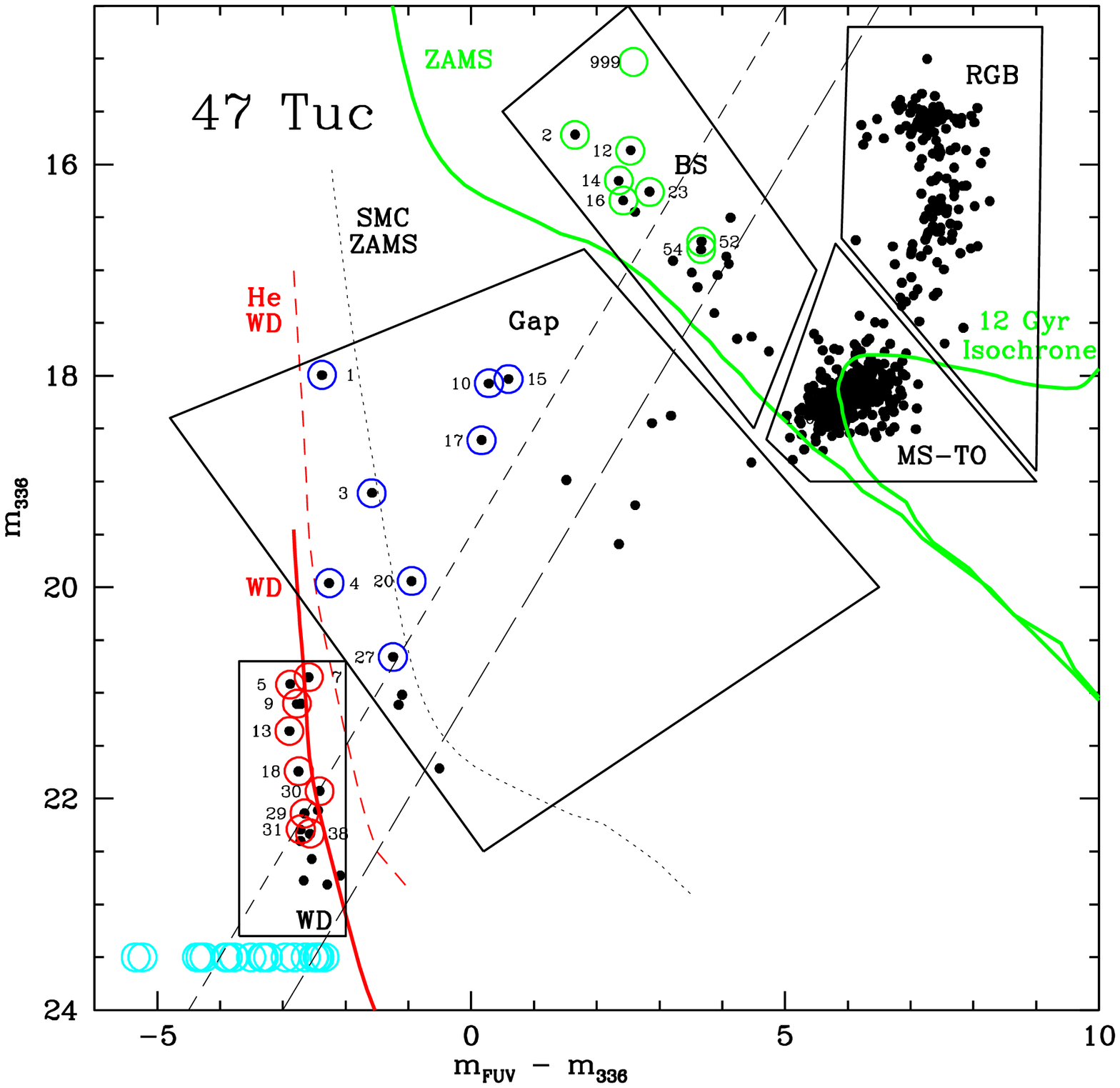}
\caption{The FUV-optical CMD of objects in the core of 47~Tuc. 
Objects detected in both the FUV and WFPC2/F336W images are shown as
black points, objects for which useful spectra could be extracted are
additionally marked with circles and labelled with their spectroscopic
ID number (c.f. Table~\protect\ref{tab:specdat}). Cyan circles
correspond to FUV sources with spectra but without optical
counterparts; all 23 of these 
are shown at $m_{F336W} = 23.5$ and without their ID numbers (to
avoid confusion). Several synthetic model sequences are shown and
labelled, as are photometric selection boxes for various
populations. We also show the location of the zero-age main sequence
in the SMC, which is located behind 47~Tuc. Note that Star~999 (in the
BS selection box) is not 
marked by a black dot since it was not in the photometric catalogue of
Paper~I (see Section~\protect\ref{sec:fuvdata}). The long-dashed diagonal
line marks the approximate spectroscopic extraction limit ($m_{FUV} = 
21.0$). The short-dashed diagonal line corresponds to $m_{FUV}
=19.5$\ and is discussed in Section~\ref{sec:binfraction}.}
\label{fig:cmd} 
\end{figure*}

Figure~\ref{fig:cmd} shows the FUV-optical CMD for sources in our
imaging field of view. The data and models shown in
Figure~\ref{fig:cmd} are exactly the same as in Paper~I, except for
(i) the slight updates to the FUV photometry described in
Section~\ref{sec:fuvdata}, (ii) the fact that we now also show
objects without optical counterparts, for which we adopt an F336W
magnitude of 23.5 (well below the actual optical detection
limit), and (iii) we have added an approximate He WD model sequence
(calculated by shifting the standard WD sequence to account for the
radius difference between a 0.5~$M_{\odot}$\ WD and
0.25~$M_{\odot}$\ one).

Objects for which we have been able to extract useful spectra are
circled in Figure~\ref{fig:cmd} and labelled with their spectroscopic
ID number (c.f. Figure~\ref{fig:images} and Table~\ref{tab:specdat}).
Each spectroscopic target is also assigned a preliminary photometric
classification, based on its position in
Figure~\ref{fig:cmd}. Possible classifications include white dwarf
(WD), ``gap object'' (sources between the WD and main sequences),
blue stragglers (BS), FUV source with no optical counterpart (NoOpt),
main-sequence turn-off star (MS-TO) and red giant/horizontal branch
star (RGB).\footnote{We speculated in Paper~I that the mismatch
between the predicted and observed position of stars on the RGB may
be due to chromospheric FUV emission from these objects. However, the
recent discovery of a significant red leak in the STIS/FUV-MAMA
response (STIS Instrument Handbook for Cycle~17; Section 5.3.4)
provides a more mundane explanation. This red leak is not yet
incorporated into the sensitivity curve used in {\sc
stsdas/synphot}. Fortunately, this has negligible impact on our
results, since all of the objects discussed in Section~\ref{sec:sed}
are relatively blue and FUV bright.}
Selection boxes for all of these objects
(except the NoOpts) are shown in Figure~\ref{fig:cmd}, and the
spectroscopic targets inside each box are marked with a distinct
colour. The photometric classifications are also given in
Table~\ref{tab:specdat}. 

%THIS IS WHERE FIG 4 GOES IN EMULATEAPJ

In the context of stellar exotica, the gap region is of particular
interest. This is the area 
in which CVs and detached WD-MS binaries may be expected to be
found (Paper~I; Townsley \& Bildsten 2002). Some such objects may also
fall into the WD and NoOpt  
selection boxes, e.g. if the hot WD still dominates the binary SED in
both FUV and F336W. However, as shown statistically in Paper~I, both
of the latter boxes are likely to contain primarily WDs. 

In total, we have been able to extract useful spectra for 48 objects,
including 8 gap sources, 8 BSs, 9 WDs and 23 NoOpts. All MS-TO and RGB 
objects were too faint to be included in our spectral extraction. The
long-dashed diagonal line corresponds to $m_{FUV} = 21.0$ and marks the
approximate spectroscopic extraction limit. Note that not all sources
above this limit are circled in Figure~\ref{fig:cmd}. This is because
some bright sources fell outside our spectroscopic field of view and
others were too severely blended with even brighter objects to yield
useful spectra. The short-dashed diagonal line corresponds to $m_{FUV}
= 19.5$ and marks the approximate spectroscopic completeness limit, as
discussed further in Section~\ref{sec:binfraction}. Using the same
method as in Paper~I, we estimate that $\simeq 1.4$\ objects in the CMD
brighter than $m_{FUV} = 21.0$, and $\simeq 0.3$ objects brighter than
$m_{FUV} = 19.5$ may be the result of a chance coincidence (i.e. a
superposition of two unrelated stars). These estimates are based on
the number of FUV sources brighter than the adopted cut-offs, the
number of optical sources, and the number of actual matches between
these source lists. 
%Amongst FUV 
%objects brighter than $m_{FUV} = 19.0$, the expected number of
%spurious coincidences is $\simeq 0.1$.

\subsection{Proper Motions and Cluster Membership}
\label{sec:proper}

\begin{figure}
%\epsscale{1.2}
\epsscale{1.0}
\plotone{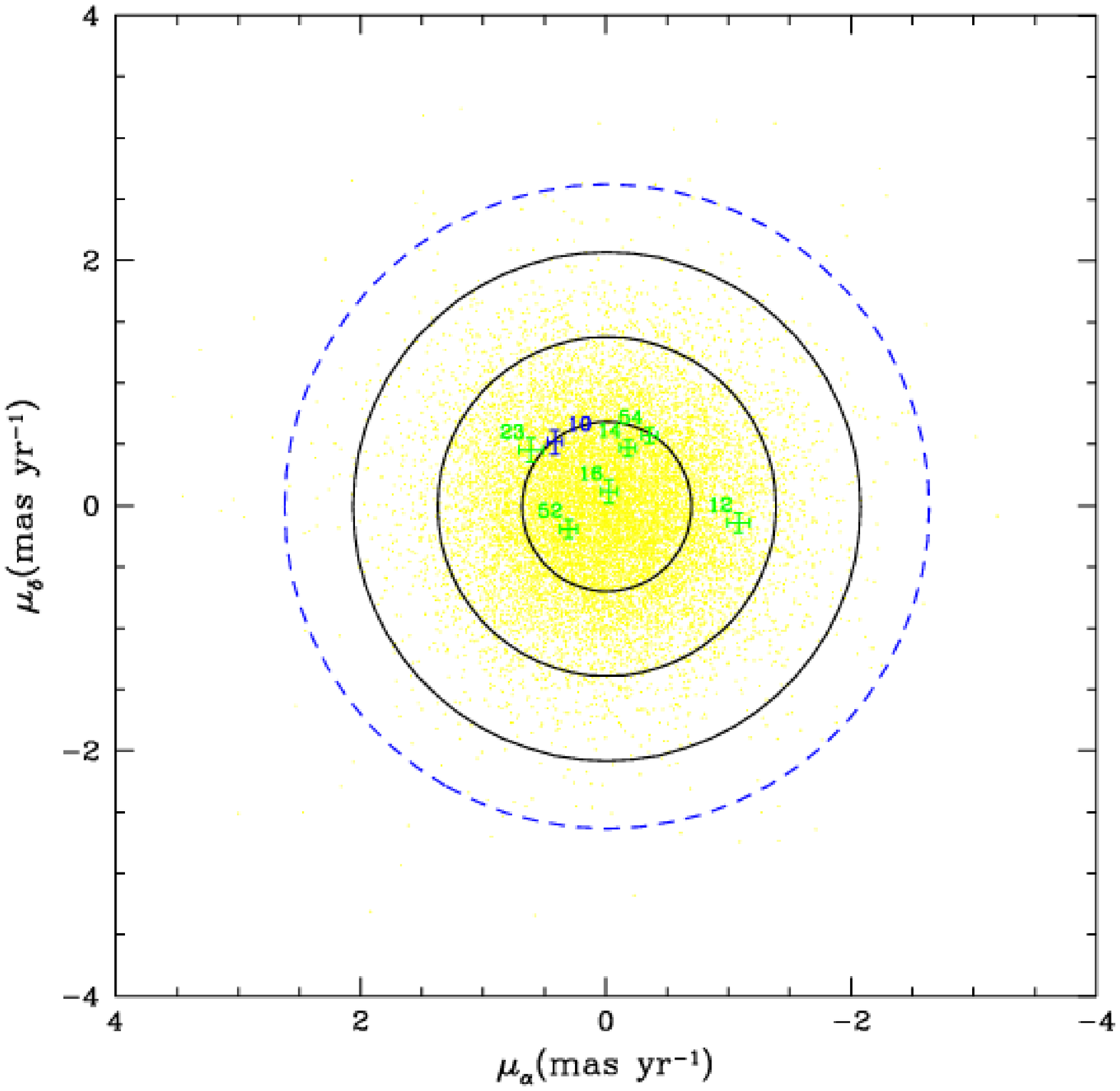}
\caption{Proper motion diagram of stars near the central regions of
  47~Tuc (constructed from data in McLaughlin et al. [2006]). The 6
  FUV sources with useful spectra for which proper motions are
  available are highlighted and labelled with their identification
  numbers. Colours correspend to the different source types, based on
  the photometric classification scheme shown in
  Figure~\protect\ref{fig:cmd} and described in
  Section~\protect\ref{sec:cmd}. Blue stragglers are shown in green,
  and the one gap object in blue. The solid black ellipses correspond
  to 1$\sigma$, 2$\sigma$, 3$\sigma$\ contours (where
  $\sigma$\ refers to the 1-dimensional standard deviations in
  $\mu_\alpha$\ and $\mu_\delta$). These contours contain 42.4\%,
  87.3\% and 98.2\% of all sources, respectively. All proper
  motions are relative to the mean motion of 47~Tuc, and the
  dashed blue circle marks the escape velocity from the cluster. Stars
  belonging to the Small Magellanic Cloud (which is located behind the
  cluster along the line of sight) would lie around $\mu_\alpha =
  -4.7$~mas~yr$^{-1}$\ and $\mu_\delta = 1.3$~mas~yr$^{-1}$ in these
  co-ordinates (beyond the right edge of the plot).}
\label{fig:proper}
\end{figure}

McLaughlin et al. (2006) have recently created a large photometric and
proper motion catalogue that contains over 14,000 stars near the
central regions of 47~Tuc. A subset of 18 of our spectroscopic targets
have counterparts in their master catalogue, and 6 of these have
well-determined proper motions. In Figure~\ref{fig:proper} we show the
location of these stars in the proper motion diagram, relative to
other cluster members in the McLaughlin catalogue. All of our
targets (which include 5 BSs and 1 gap object according to the
CMD-based classification) are fully consistent with being cluster
members.

% THIS IS WHERE F5 GOES IN EMULATEAPJ

\section{FUV Spectroscopy of 48 Objects in the Core of 47 Tuc}
\label{sec:overview}

In this section, we present an overview of all the useful spectra 
we could extract from our slitless FUV spectral images. We have
grouped the spectra by their CMD-based classifications for this
purpose.

\subsection{The Spectra of Gap Objects}
\label{sec:gap_specs}

\begin{figure*} 
\center
\includegraphics[angle=-90,width=1.0\textwidth,viewport=40 40 520 730,clip]{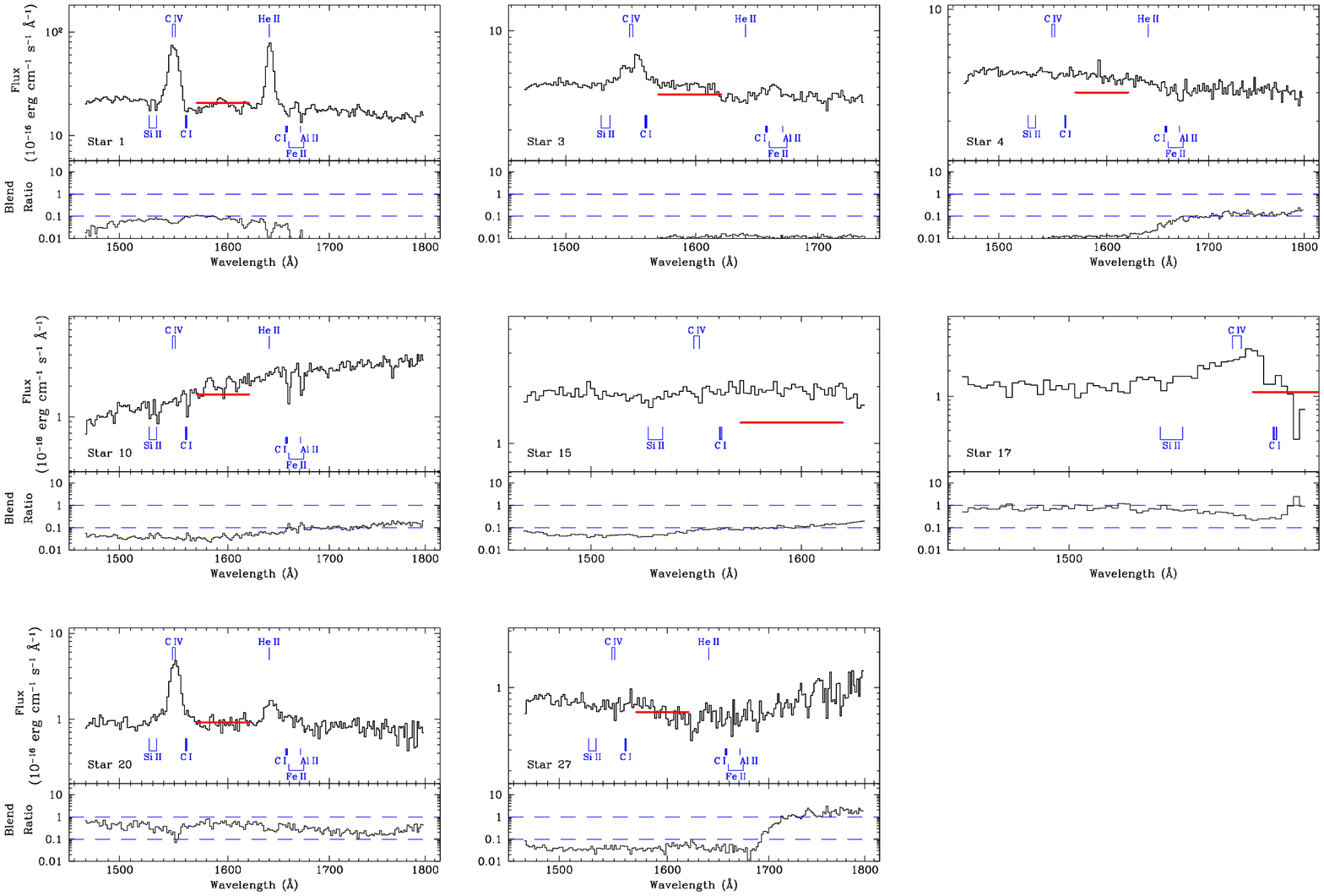}
\caption{\scriptsize The FUV spectra of gap objects. The top panel for each object
  shows the extracted spectrum. The positions and identifications of
  several key spectral lines are also indicated. The red horizontal
  line marks the flux level corresponding to the FUV photometry (for a
  flat spectrum source; see Equation~\ref{eq:stmag} in
  Section~\protect\ref{sec:optdata}). Note that the wavelength range
  covered for each object depends on its position on the detector. 
  The bottom panel for each
  objects shows the wavelength-dependent blend ratio, as defined in
  Section~\protect\ref{sec:fuvdata}. The horizontal dashed lines mark
  blend ratios of 0.1 and 1.0; this is the range where blending can start
  to become a problem.}
\label{fig:gap_specs}
\end{figure*}

The FUV spectra of our 8 gap sources are shown in
Figure~\ref{fig:gap_specs}. The diversity of these spectra is striking
and immediately suggests that these objects do not form a homogenous
group. All gap sources will be analyzed individually in
Section~\ref{sec:sed}, so we only present a brief overview
here.

% THIS IS WHERE F6 GOES IN EMULATEAPJ

% THIS IS WHERE F7 GOES IN EMULATEAPJ

Clear emission lines are seen in Stars~1, 3, and 20, all of which were
previously known or suspected CVs (also known as AKO~9, 
V1, and V2, respectively [see Table~\ref{tab:specdat}]). The presence
of emission lines confirms the CVs nature of these objects. Our
data on AKO~9 has already been analyzed in detail elsewhere (Knigge et
al. 2003), but the spectroscopic confirmation of V1 and V2 as CVs is
reported here for the first time in the refereed literature.
\footnote{The review presented in Knigge (2004) inludes preliminary,
  rough spectral extractions for these objects.}
Only one other gap source -- Star~17 -- shows tentative evidence of
line emission. Unfortunately, the position of this object on the
detector is such that only a short segment of its spectrum is
available, with the red cut-off occuring just beyond the C~{\sc iv}
line. The line itself thus falls close to the edge of the detector,
which is why we do not consider the apparent flux excess near
1550~\AA\ to be a definite detection of line emission. We will discuss
Star~17 in more detail in Section~\ref{sec:star17}. 

Of the 4 remaining objects, 3 exhibit flat or blue continua, but the
fourth -- Star~10 -- presents a rather strange, red continuum with
absorption lines. All 4 of these objects turn out to be interesting 
(see Section~\ref{sec:sed}), but Star~10, in particular, is
probably the most exotic and unusual object our FUV survey has
uncovered. This object is analyzed more closely in
Section~\ref{sec:star10}.

\subsection{The Spectra of Blue Stragglers}
\label{sec:bs_specs}

\begin{figure*} 
\center
\includegraphics[angle=-90,width=1.0\textwidth,viewport=40 40 520 730,clip]{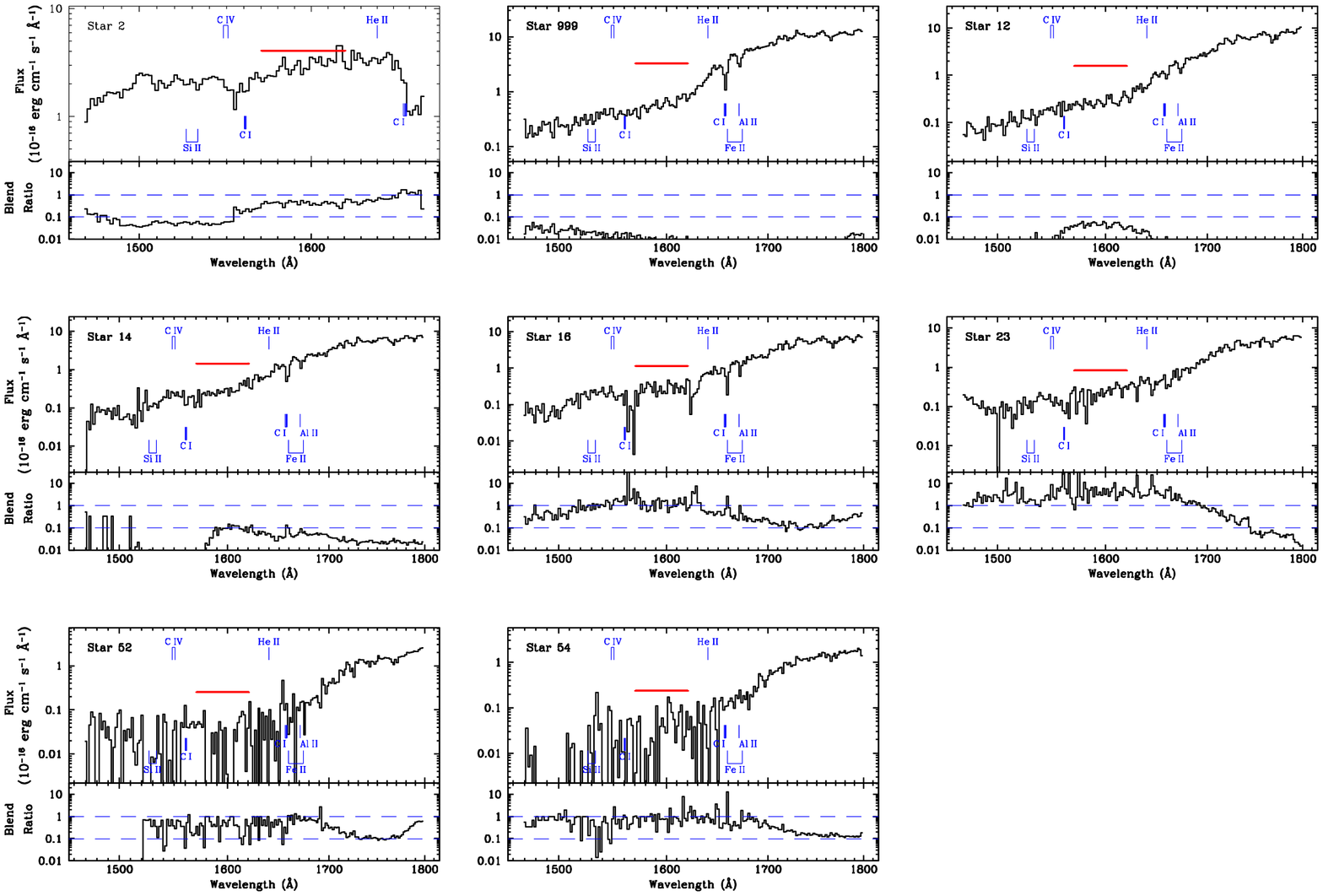}
\caption{The FUV spectra of blue stragglers. See caption to
  Figure~\protect\ref{fig:gap_specs} for details.}   
\label{fig:bs_specs}
\end{figure*}

Figure~\ref{fig:bs_specs} shows the FUV spectra of our 8 BSs in the
core of 47~Tuc. Only the two brightest of these -- Stars~2 and 999 --
will be analyzed in more detail in Section~\ref{sec:sed}.

Unlike the gap sources, all but one of the BSs display rather similar
FUV spectra. More specifically, 7 of the 8 BSs show extremely red FUV
continua, with fluxes increasing by typically 2 orders of magnitude
between roughly 1500~\AA\ and 1800~\AA. This near-exponential rise is
expected, since BSs are stars of spectral type A-F for which the FUV waveband
falls on the Wien tail of their SEDs. In principle, this should make
the FUV flux and spectral shape a very sensitive thermometer for BSs, 
a possibility we will explore in Section~\ref{sec:star999}. In those
sources with sufficient S/N, the spectra also exhibit absorption lines
due to C~{\sc i} and Al~{\sc ii} (with a possible contribution from
Fe~{\sc ii}).

The obvious outlier among the BSs is Star~2, whose FUV continuum rises
much more slowly towards longer wavelengths. Moreover, the flux level
at the blue end of the spectrum (around 1500~\AA) is much higher than
for any other BS in our sample. Thus even though the red part of
Star~2's FUV spectrum may be affected by blending, the discrepancy 
between this object and the other BSs appears to be real. Star~2 is
analyzed more carefully in Section~\ref{sec:star2}. 

\subsection{The Spectra of White Dwarfs}
\label{sec:wd_specs}

\begin{figure*} 
\center
\includegraphics[angle=-90,width=1.0\textwidth,viewport=40 40 520  730,clip]{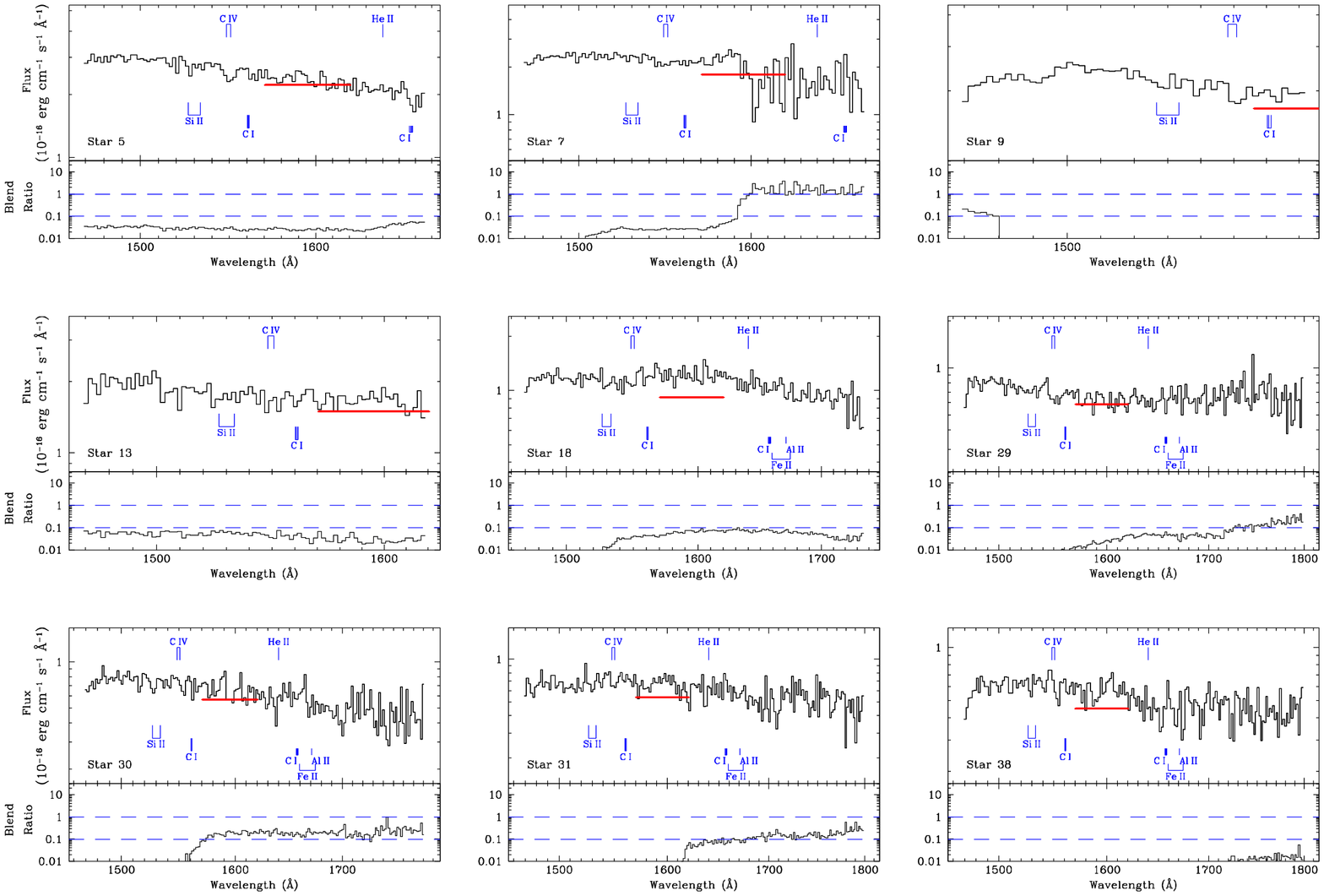}
\caption{The FUV spectra of hot white dwarfs. See caption to
  Figure~\protect\ref{fig:gap_specs} for details.}   
\label{fig:wd_specs}
\end{figure*}

The FUV spectra of the 9 hot WDs in our sample are shown in
Figure~\ref{fig:wd_specs}. Only the two brightest of these -- Stars~5
and 7 -- will be analyzed in more detail in
Section~\ref{sec:sed}. 

The WDs present a relatively homogenous set of FUV spectra. which can
be characterized as blue, featureless continua. This is as expected
for young WDs near the top of the cooling sequence. There is
little evidence for absorption or emission lines in any of these
spectra. The only features that are perhaps worth remarking
on are a hint of an absorption dip near C~{\sc iv} in the spectrum of
Star~5, and a hint of emission near the same line in the spectrum of
Star~29. However, we regard neither of these features as convincing. We
thus conclude that the top of the WD cooling sequence in 47~Tuc
does not hide a large population of WD-dominated CVs (see also
Section~\ref{sec:no_specs}). Surprises are still possible though, as
illustrated by the analysis of Star~7 in Section~\ref{sec:star7}.

% THIS IS WHERE F8 GOES IN EMULATEAPJ

\subsection{The Spectra of FUV Sources without Optical Counterparts}
\label{sec:no_specs}

\begin{figure*} 
\center
\includegraphics[angle=-90,width=0.95\textwidth,viewport=40 40 520 730,clip]{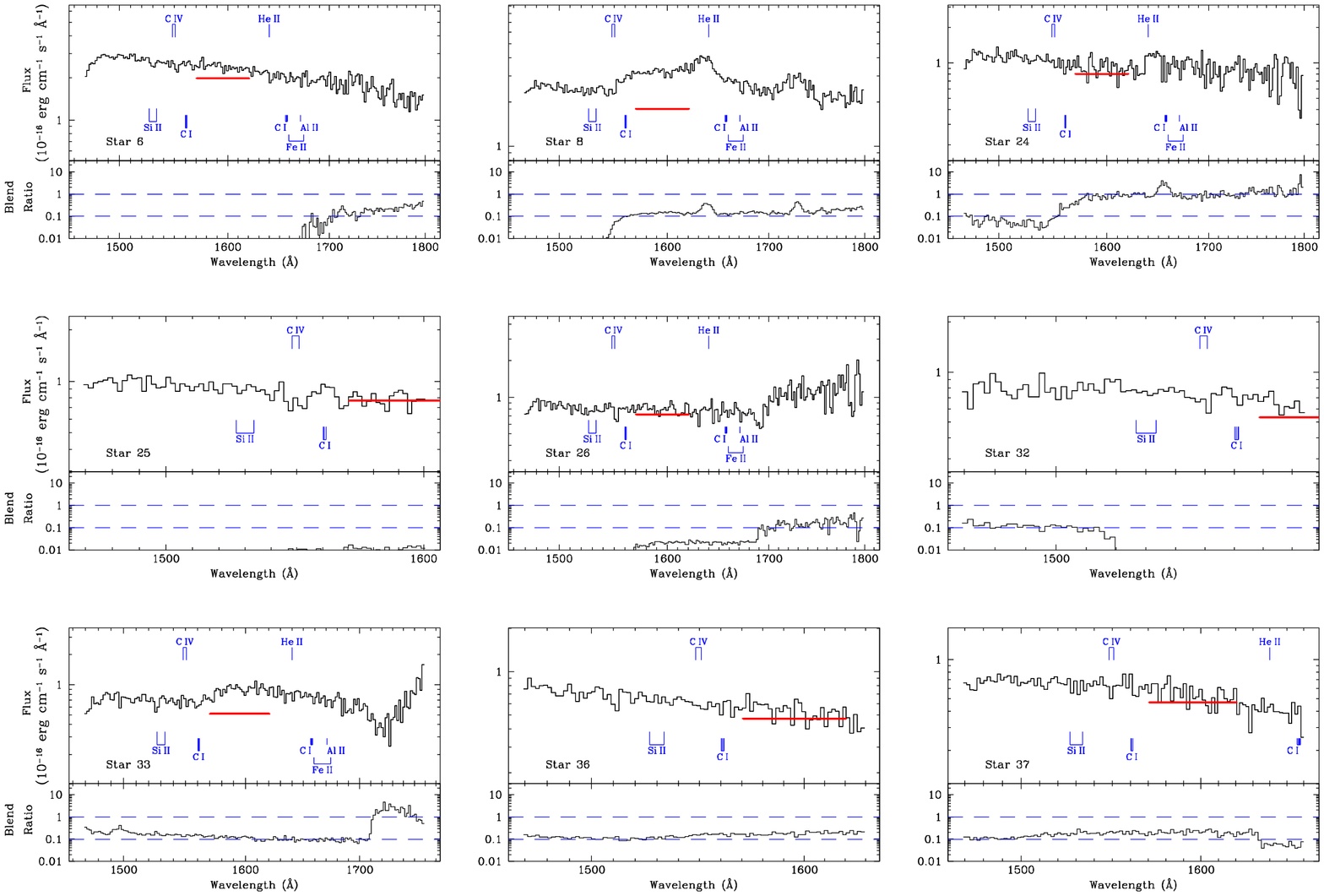}
\includegraphics[angle=-90,width=0.95\textwidth,viewport=40 40 520 730,clip]{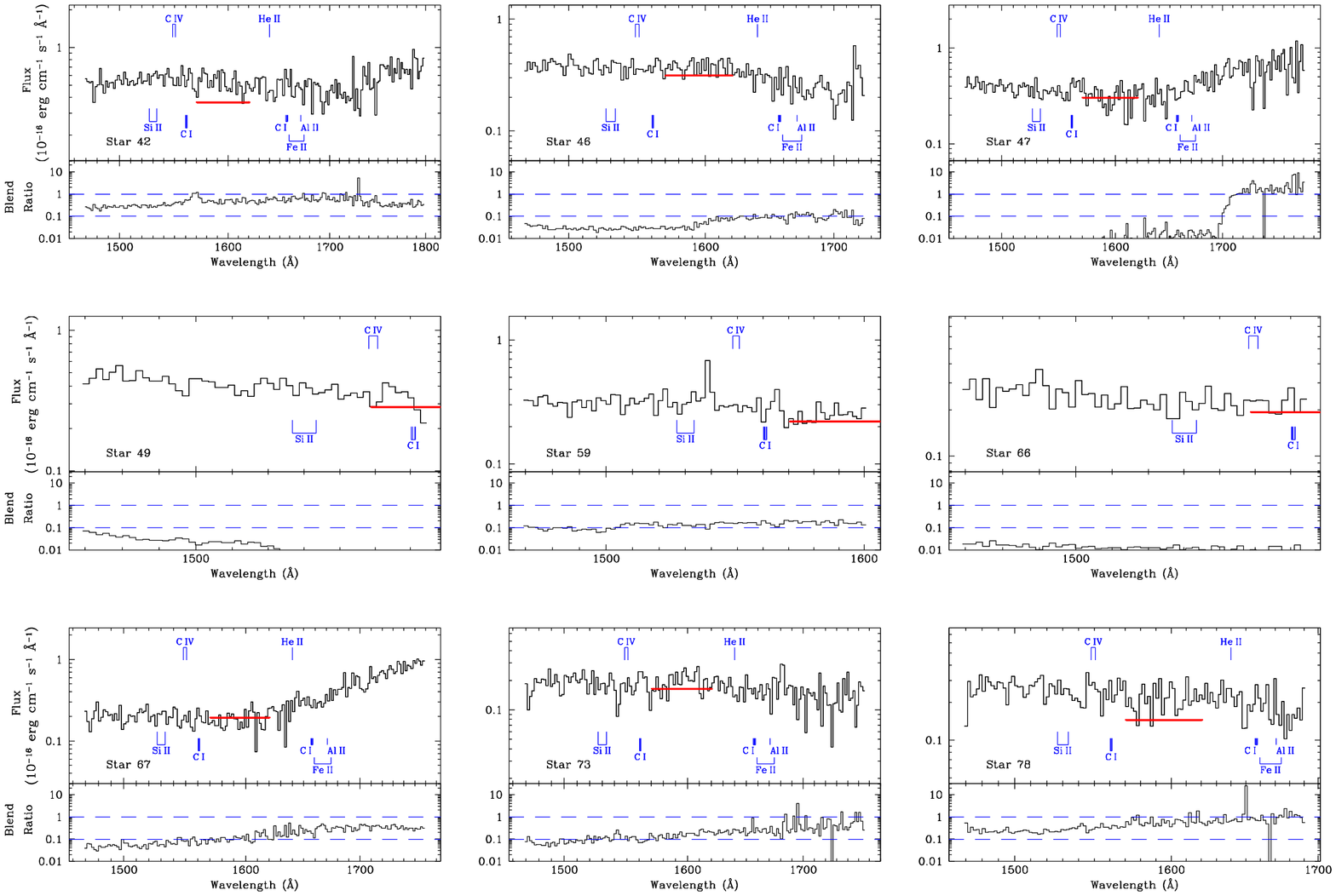}
\caption{The FUV spectra of objects without optical counterparts. See caption to
  Figure~\protect\ref{fig:gap_specs} for details.}   
\label{fig:no_specs}
\end{figure*}

Figure~\ref{fig:no_specs} shows the spectra of our 23 FUV sources 
without optical counterparts. This group of objects is also expected
to be dominated by hot WDs. As shown in Paper~I, the number of these
objects is consistent with expectations for the WD population, and our
optical (WFPC2/PC/F336W) photometry is certainly not complete for
these optically faint objects. Since we do not have optical
counterparts, none of these sources are included in
Section~\ref{sec:sed}.

\begin{figure*}[t] 
\center
\figurenum{8 (cont)}
\includegraphics[angle=-90,width=0.95\textwidth,viewport=40 40 360 730,clip]{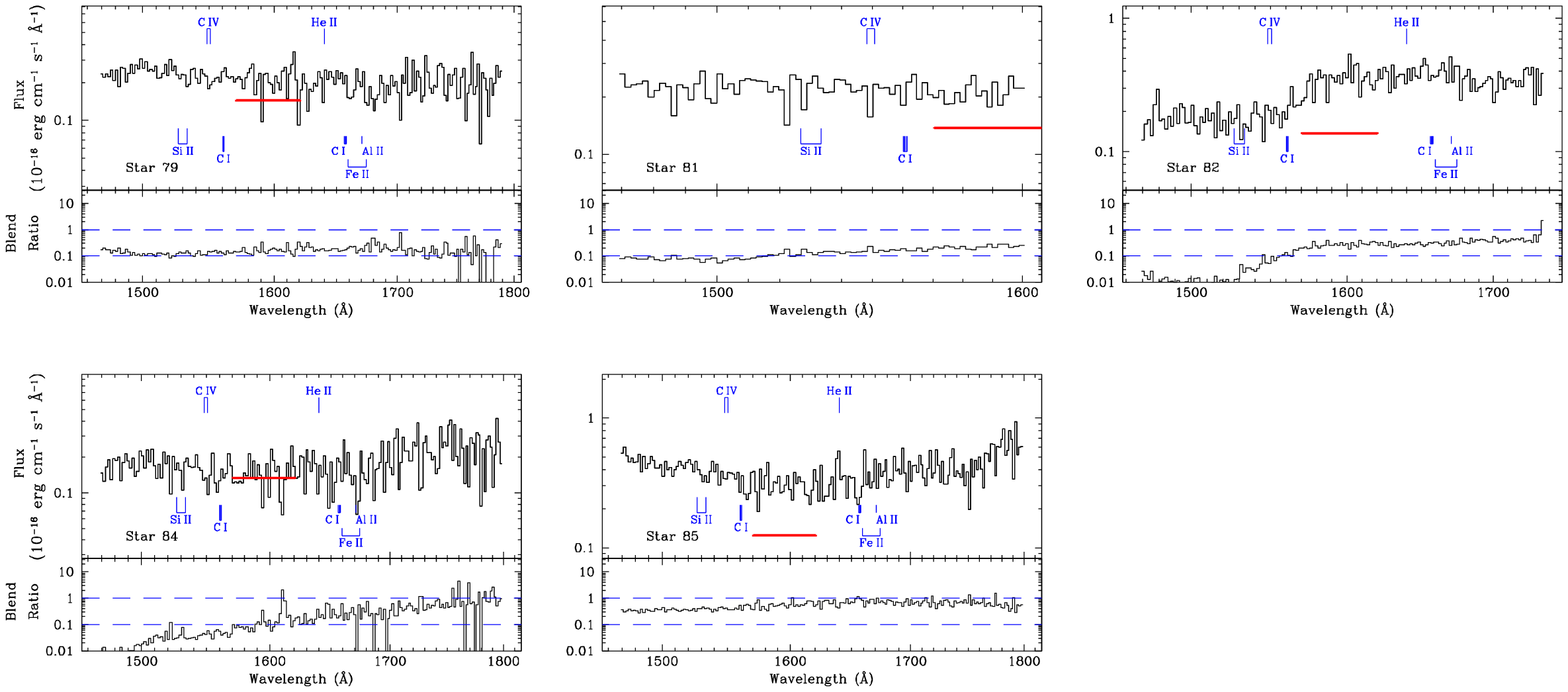}
\caption{The FUV spectra of objects without optical counterparts.}
\end{figure*}

Figure~\ref{fig:no_specs} shows that the spectra of essentially all of
these FUV sources are consistent with expectations for single DA
WDs. In particular, the resemblance of the spectra to those of the
photometrically-classified WDs (Figure~\ref{fig:wd_specs}) is
obvious. The spectrum of Star~33 contains an obvious hump, but this is
caused by blending with a fainter FUV source. That source was not
included in the spectral extraction, which explains why the blend
ratio for Star~33 does not show this feature. None of the objects in
Figure~\ref{fig:no_specs} show convincing evidence for emission or
absorption lines. This confirms the conclusion of Section~\ref{sec:wd_specs}
that the top of the cooling sequence does not harbour numerous CVs.

\section{A Closer Look at Individual Sources}
\label{sec:sed}

We will now present a more detailed analysis of the most interesting
and representative sources in our spectroscopic sample. This includes
all of the gap sources, as well as the two brightest objects from the
CMD-based WD and BS groups. All sources analysed in this section are
listed in bold at the top of Table~\ref{tab:specdat}. The last column
in the table presents our final classification for each source. 

Throughout this section, we will use model fits to the combined
spectral and SED data sets to shed light on the physical nature of our
sources. The cluster parameters adopted in these models are those of
Gratton et al. (2003), i.e. $d = 4840$~pc, $E(B-V) = 0.024$, $[Fe/H] =
-0.66$. All stars, except WDs, are described by interpolating on the
latest, $\alpha$-enhanced Kurucz {\sc atlas9} model atmosphere
grids.
\footnote{The Kurucz model atmosphere grids are available for download
  at http://kurucz.harvard.edu/grids.html.} 
These grids use the updated opacity distribution functions
described by Castelli \& Kurucz (2001) and account for enhanced
$\alpha$-element abundances at a level of $[\alpha/Fe] = 0.4$. For the
purpose of modelling WDs, we rely on the grid of synthetic WD spectra
described in G\"ansicke, Beuermann \& de Martino (1995), which has
been kindly provided by Boris G\"ansicke. We generally adopt $M_{WD} =
0.5~M_{\odot}$\ and $R_{wd} = 0.017~R_{\odot}$\ (corresponding to
$log{\;g} = 7.6$) to describe WDs in 47~Tuc. This is appropriate for
recently formed WDs near the top of the cooling sequence (Renzini \&
Fusi Pecci 1988; Wood 1995). We also use the WD models
to qualitatively describe other hot spectral components, such as
accretion disks in CVs. We will only be interested in rough estimates
of the characteristic temperatures and sizes of such generic hot
components, and the use of WD models should be adequate for this
purpose. 

The spectral resolution of the Kurucz models is relatively low, with a
typical wavelength step of 10~\AA\ in the FUV region. This turns out to
be quite adequate for our qualitative modelling, and so the observed
spectra (and also the WD models) are interpolated onto the wavelength
grid of the Kurucz spectra during our fitting procedure. We have also
carried out synthetic photometry on all models in the Kurucz and WD
grids. This was done using {\sc synspec} in {\sc iraf/stsdas}, which 
includes throughput files for all {\em HST} detector/filter
combinations. Most of our modelling is done in a least squares sense,
with every data point being given equal weight. Since we only consider
the coarse Kurucz wavelength grid when modelling the FUV spectrum, this
usually provides a reasonable compromise between the emphasis placed
on the spectrum relative to the broad-band SED. In a few cases, the
relative weights assigned to the spectroscopy and photometry were
adjusted slightly to improve the overall quality of the fits.

We note from the outset that our goal here is to find physically
plausible descriptions of the data, rather than arbitrary sets of 
best-fit parameters. Thus we will sometimes constrain certain
parameters or parameter combinations to have values consistent with,
for example, the expected location of the WD or main sequences. Of
course, we will only do this when adequate fits can actually be found
with the constrained parameters. We feel this approach is appropriate,
since the match between data and models is limited by systematic
effects (e.g. blending in the spectra, crowding in the photometry, and
uncertainties affecting the theoretical models) rather than by purely
statistical errors on the data. 

\subsection{Star~1 = AKO~9: A CV with a Subgiant Donor}
\label{sec:star1}

\begin{figure*} 
\center
\includegraphics[angle=-90,width=1.0\textwidth]{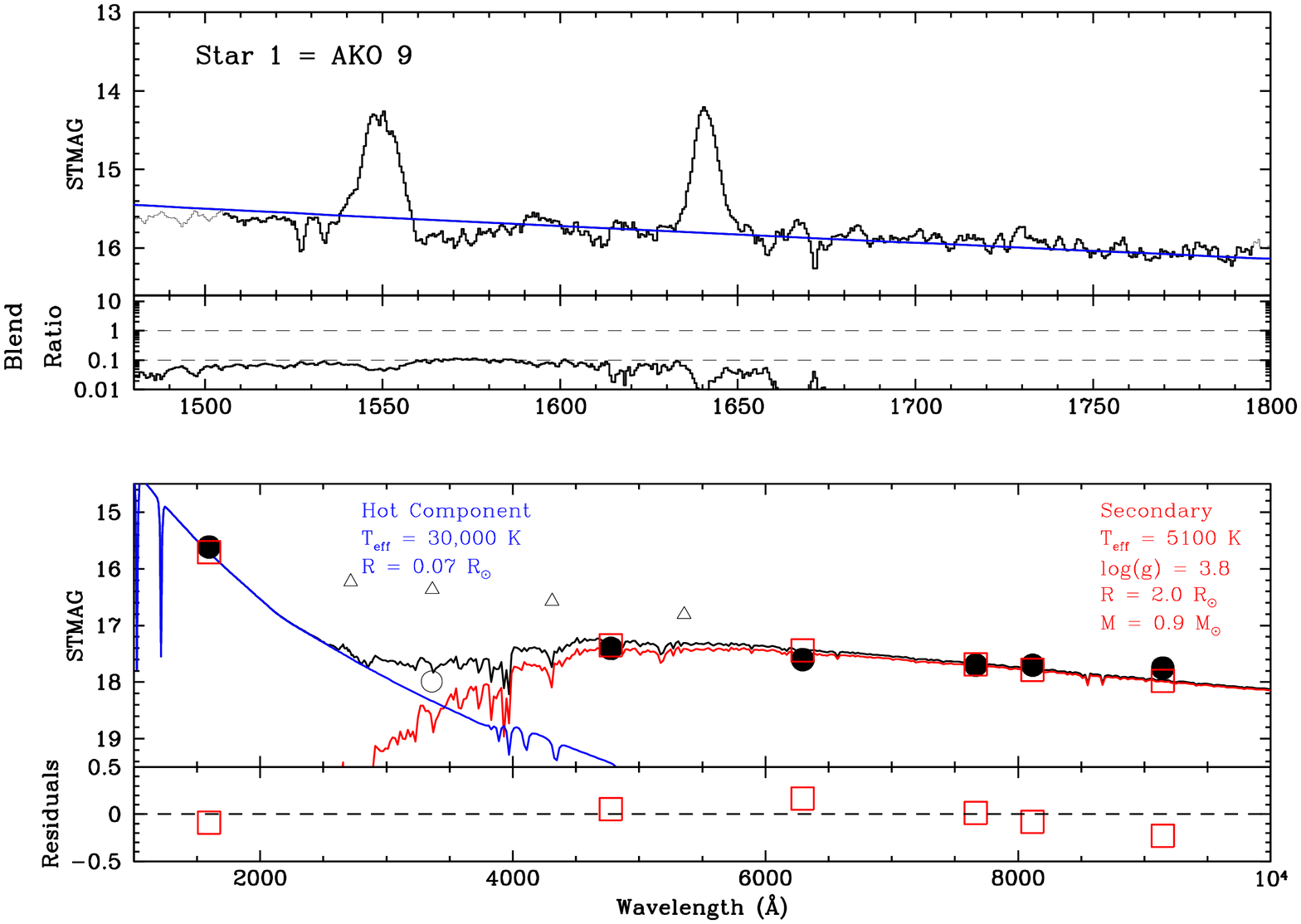}
\caption{
FUV spectrum (top panel) and broad-band SED (bottom panel) of Star~1 =
AKO~9, along with a two-component model fit to these observations. The
black histogram in the top panel is the observed FUV spectrum (shown
at the full resolution), while the blue, red and black lines (in both
panels) show the
model spectra of the hot, cool and combined components,
respectively. Only the wavelength regions shown with a thick solid line
were used in the spectral fit. Underneath the spectrum, we also show
the corresponding blend ratio, as defined in 
Section~\protect\ref{sec:fuvdata}.  
In the bottom panel, the filled black circles are 
the ACS/HRC photometric data we try to fit, whereas the open red
squares are the 
magnitudes predicted by the full model. The corresponding residuals
are also shown. The emission lines in the FUV spectrum
confirm this object as a CV, while the broad-band SED
establishes the evolved nature of its mass-losing secondary. The open
circle around 3400~\AA\ in the bottom panel is our older 
WFPC2/PC/F336W data point, and the open triangles are the ACS/HRC data
points obtained in July 2002. The system was caught in a dwarf nova
outburst during the 2002 observations (see also
Figure~\protect\ref{fig:star1phot}).}  
\label{fig:star1}
\end{figure*}

\begin{figure*}[h!]
\center
\includegraphics[angle=0,width=0.95\textwidth]{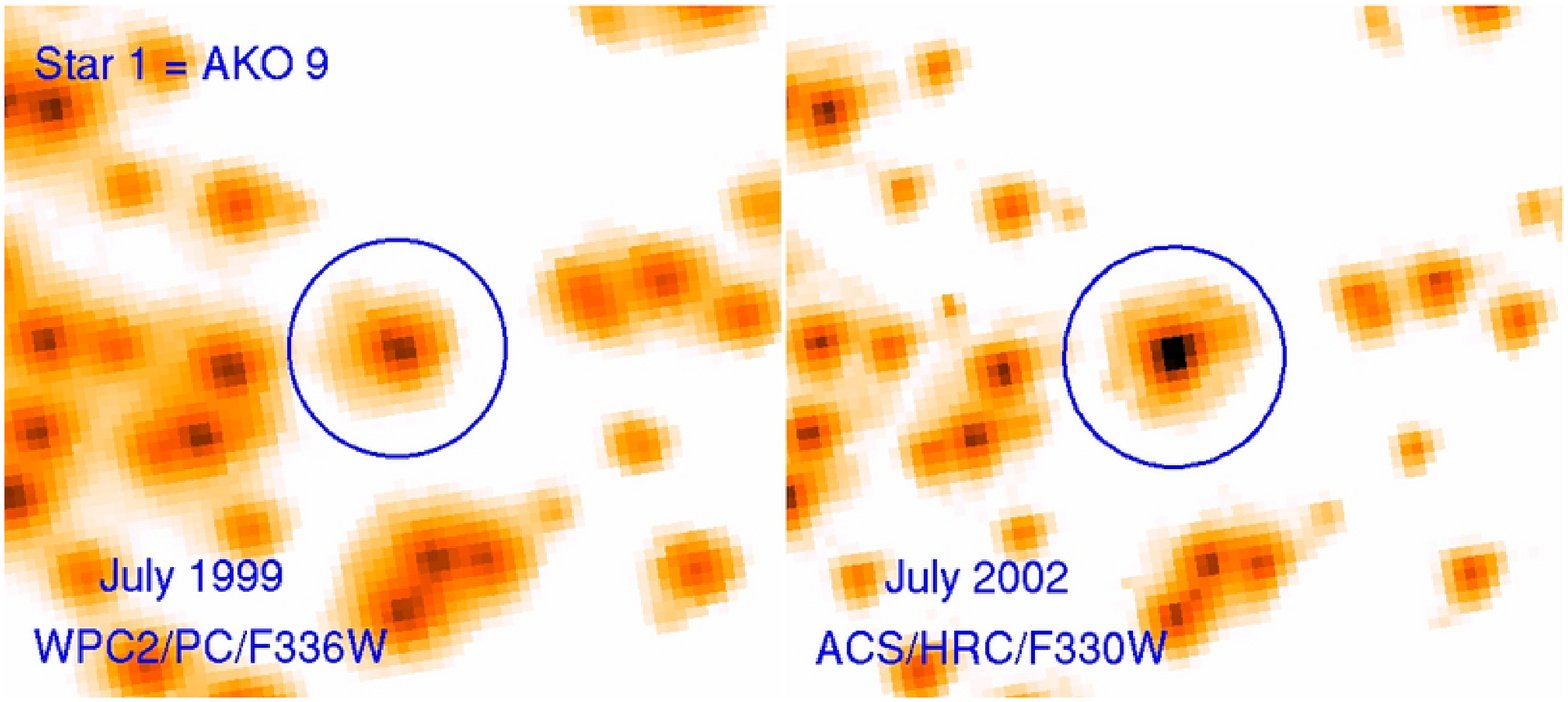}
\caption{Approximately U-band observations of AKO~9 obtained with {\em HST}
in July 1999 (left panel) and July 2002 (right panel). The difference in
brightness is obvious and implies that AKO~9 experienced a dwarf nova
eruption in July 2002. In these images, north is up, east is to the left, 
and the field of view is approximately 1.8\arcsec.} 
\label{fig:star1phot}
\end{figure*}

Star~1 is the brightest FUV source in 47~Tuc and is identical to the
previously known blue object AKO~9. A detailed analysis of our FUV
spectroscopy and photometry for this source has already been presented
in Knigge et al. (2003). Figure~\ref{fig:star1} shows the FUV spectrum
and broad-band SED for AKO~9, along with our suggested two-component
fit to these data. 

As shown in Knigge et al. (2003), AKO~9 is almost certainly a
CV with an orbital period of $P_{orb} = 1.1091$~d
and an evolved, probably subgiant donor star. This description also
turns out to provide a good fit to the more extensive broad-band
SED we have assembled here. In constructing this model, we have
assumed that the secondary is a turn-off mass object and have used the
orbital period-mean density relation for Roche-lobe filling stars,  
\begin{equation}
\left< \rho \right> \simeq 107 P_{orb,hrs}^{-2} {\rm g~cm^{-3}},
\label{eq:rho}
\end{equation}
as an additional constraint on the donor parameters (e.g. Warner
1995). The hot component  
that dominates the FUV light is too bright to be a pure WD and is
instead likely to arise in an accretion disk. In our fit, we have
described this component with a $\log{~g} = 7$~WD model atmosphere. The
temperature was fixed at $T_{hot} = 30,000$~K, which yields an effective
radius of $R_{hot} \simeq 0.07~R_{\odot}$ (for a pure WD, we would
expect $R_{hot} \simeq 0.02~R_{\odot}$). However, the temperature and
effective radius are highly correlated and not well constrained. 

In the process of constructing the broad-band SED, we noticed that a
subset of optical observations obtained in July 2002 were
systematically brighter than all of our other optical data. We show
these 4 data points as open triangles in Figure~\ref{fig:star1}. Note,
in particular, the almost 2~mag brightness difference between the
roughly U-band data used in Paper~I (open circle in
Figure~\ref{fig:star1}) and that obtained in July 2002. As shown in
Figure~\ref{fig:star1phot}, this difference is easily noticeable in
the images themselves.

We conclude that we have discovered a dwarf nova eruption of
AKO~9. Two previous outburst of this object have been observed with
{\em HST}, one in July 1986, the other in October 1992 (Minniti et al. 1997;
see also Knigge al. al. 2003). The present discovery of an eruption in
July 2002 confirms that the mean interoutburst recurrence time is at
most 6-7 years. However, given the relatively sparse time coverage of
the photometry investigated to date, a shorter recurrence time remains
possible.

Even though all of the existing observational evidence is consistent
with AKO~9 being a dwarf nova-type CV, the source warrants additional
investigation. In particular, a radial velocity study would be
invaluable for establishing the system parameters. As an incentive for
such a study, we note that the X-ray, FUV and optical properties
established so far do not completely preclude the possibility that the
primary could be a black hole, rather than a WD.

\subsection{Star~2: A Blue Straggler with a White Dwarf Companion}
\label{sec:star2}

\begin{figure*} 
\center
\includegraphics[angle=-90,width=1.0\textwidth]{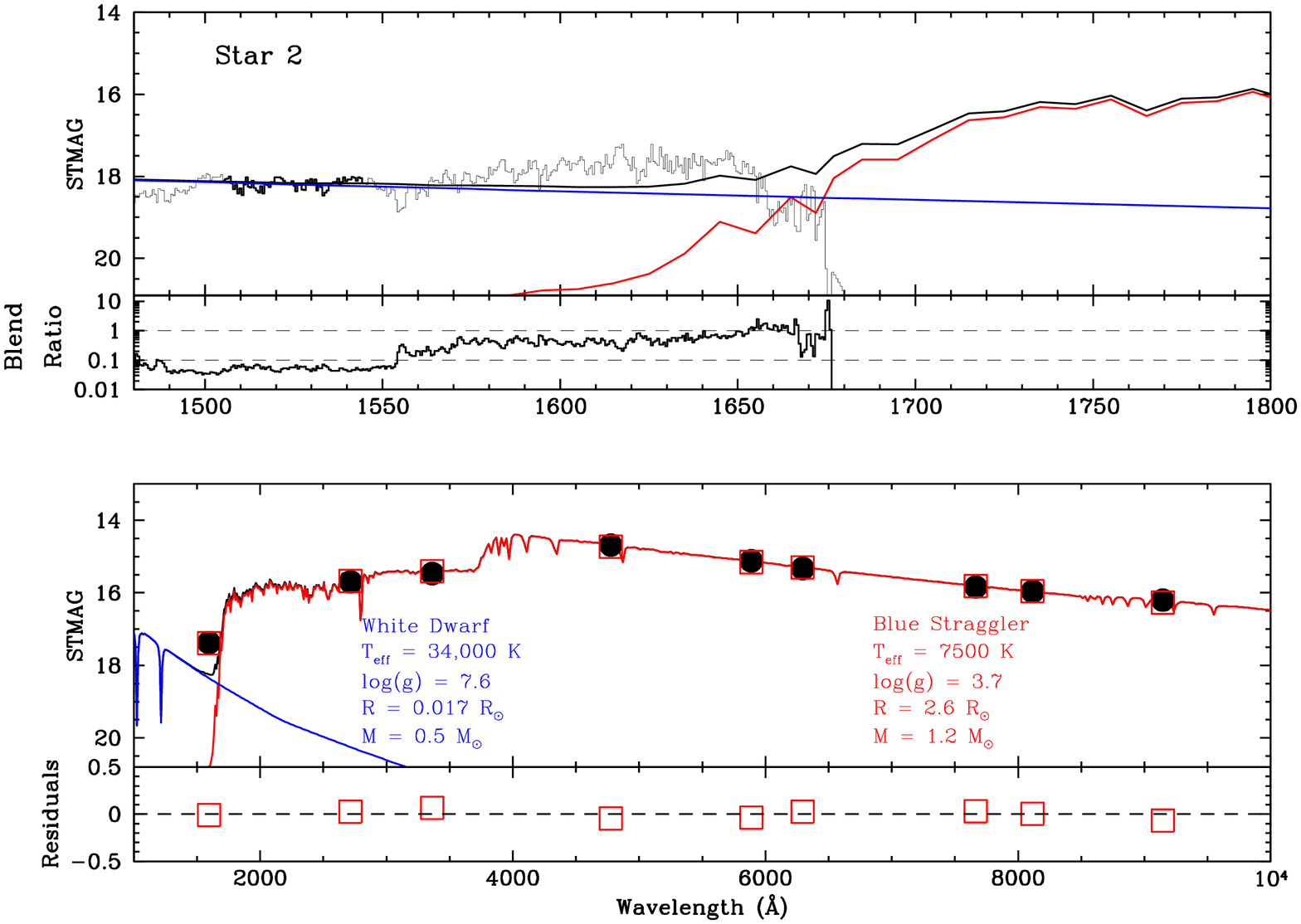}
\caption{The FUV spectrum (top panel) and broad-band SED (bottom
panel) of Star~2, along with our best-bet model fit to the
data (see text for details). Notation is as in
Figure~\protect\ref{fig:star1}. The region of the spectrum beyond 1550
\AA\ was excluded from the fit because of blending with other objects, 
but there is clear evidence for strong, unblended emission shortward
of this. Star~2 appears to be a blue straggler with a hot WD companion.}
\label{fig:star2}
\end{figure*}

Star~2 lies on the BS sequence in the FUV-optical CMD
(Figure~\ref{fig:cmd}), but its FUV spectrum rises much more slowly
towards longer wavelengths than that of any other BS in our data
(Figure~\ref{fig:bs_specs}). Even though our spectrum for this object
is rather short, and may be affected by blending longwards of about
1550~\AA, we have no reason to doubt the reality of the FUV excess it
implies (relative to other BSs). 

Figure~\ref{fig:star2} shows the FUV spectrum and broad-band SED for
this object, along with our best-bet model. All of the photometry is
well described by a single component with parameters appropriate for a
BS in 47 Tuc. However, no single component model is able to also match
the FUV spectrosopy.
\footnote{The surface gravity (and hence mass) of the BS is not well 
constrained by the fit and was therefore fixed at a physically
plausible value. Also, the predicted FUV spectrum of the BS may not be
very accurate (see Section~\ref{sec:star999}). However, our conclusion
that the observed FUV spectrum shows a strong excess at the shortest
wavelengths is insensitive to these caveats.}

The simplest and most plausible way to account for the observed FUV
excess is to invoke a WD companion to the BS. As shown in 
Figure~\ref{fig:star2}, a WD with $T_{eff} \simeq 34,000$~K provides a
reasonable match to the FUV spectrum without significantly affecting 
the broad-band SED. Thus a BS-WD binary system provides a good
match to the entire spectroscopic and potometric data set.

Is it possible that we are being fooled by a chance superposition?
There are two questions to consider in this context. First, could
the FUV flux we have identified with the short-wavelength spectrum of
Star~2 actually be 
associated with a different FUV source? Inspection of the direct image
(Figure~\ref{fig:images}) shows that there are two FUV-bright 
sources -- Stars~17 and 30 -- that lie close enough to Star~2 in 
the spatial direction to warrant consideration. However, Star~17
lies to the right of Star~2 in the direct 
image and so cannot be responsible for any short-wavelength flux in
the spectrum of Star~2. Star~30, on the other hand, has photometric 
and spectroscopic properties consistent with a single WD and lies
about 585 pixels (corresponding to about 340~\AA) to the left of
Star~2. If the short-wavelength flux attributed to Star~2 were really
due to Star~30, essentially of the observed counts would have to be
associated with wavelengths well beyond 1800~\AA. However, the FUV
bandpass has very little sensitivity at such long wavelengths, and the
spectroscopic countrates of even the reddest FUV-bright sources in our
sample -- the BSs -- peak well shortward of 1800~\AA. Thus Star~30 is
almost certainly not the source of any flux assigned to Star~2.

The second question is whether the FUV source we see in the direct
image could be a chance blend of the bright BS with a comparably
bright WD. It is worth noting here that the BS is {\em expected} to be 
detected in the FUV, so it is not plausible that the WD should 
completely dominate the FUV output. We have inspected 
the direct FUV image in the vicinity of Star~2 and find no evidence to
support the idea that the FUV source is a blend. However, it should be
acknowledged that the FUV imaging PSF 
is rather complex and asymmetric, and that there are not many
FUV-bright objects against which Star~2 could be meaningfully
compared. By the same token, however, the 
probability of two such bright objects lying close to each other by 
chance is very small. Figure~\ref{fig:star2} shows that both the WD
and the BS would be brighter than $m_{FUV} \simeq 19$ on their
own. In total, there are only 20 objects (including Star~2) that are
this bright in our FUV photometry. The probability that two such
objects should lie close enough to each other to form an
unrecognizable blend (say within 3~pixels) is only $\simeq 1$\%. Thus
it is highly likely that the hot WD is physically associated with the
BS.

Even though Star~2 is the first BS-WD binary that has been identified
in any GC to date, the existence of such objects should probably be
expected. One straightforward way to form such a system
is directly from a MS-MS binary. As the more massive object in this
progenitor system ascends the giant branch, it may overflow its Roche
lobe and thus turn its companion into a BS via Case~B or Case~C mass
transfer
\footnote{Here, Case~B refers to mass transfer occuring prior to He
ignition (during the ascent of the red giant branch), which may
produce a He WD; Case~C refers to mass transfer after He ignition
(during the ascent of the asymptotic giant branch), which would leave 
behind a CO WD (e.g. Paczy{\'n}ski 1971).}

In either case, the outcome would be a BS-WD
binary. This formation channel will be available in GC cores as long 
as the original binary is hard (since otherwise it would not survive
until the onset of mass transfer). The hard/soft boundary in GCs
corresponds to binary separations on the order of a few AU
(e.g. Davies 1997), so there is certainly room for BS-WD progenitor
systems in the parameter space of hard binaries. 

An alternative way to form BS-WD binaries is directly via dynamical
encounters. In particular, since single BSs are the most massive stars 
found in GCs, any 3-body interaction involving such a BS would 
tend to leave it with a companion (usually the more massive member of
the original binary). Thus a system like Star~2 could have been
formed, for example, in an exchange encounter involving a BS and a
WD-MS binary. 

\subsection{Star~3 = V1: A CV with a Main-Sequence Donor}
\label{sec:star3}

\begin{figure*} 
\center
\includegraphics[angle=-90,width=1.0\textwidth]{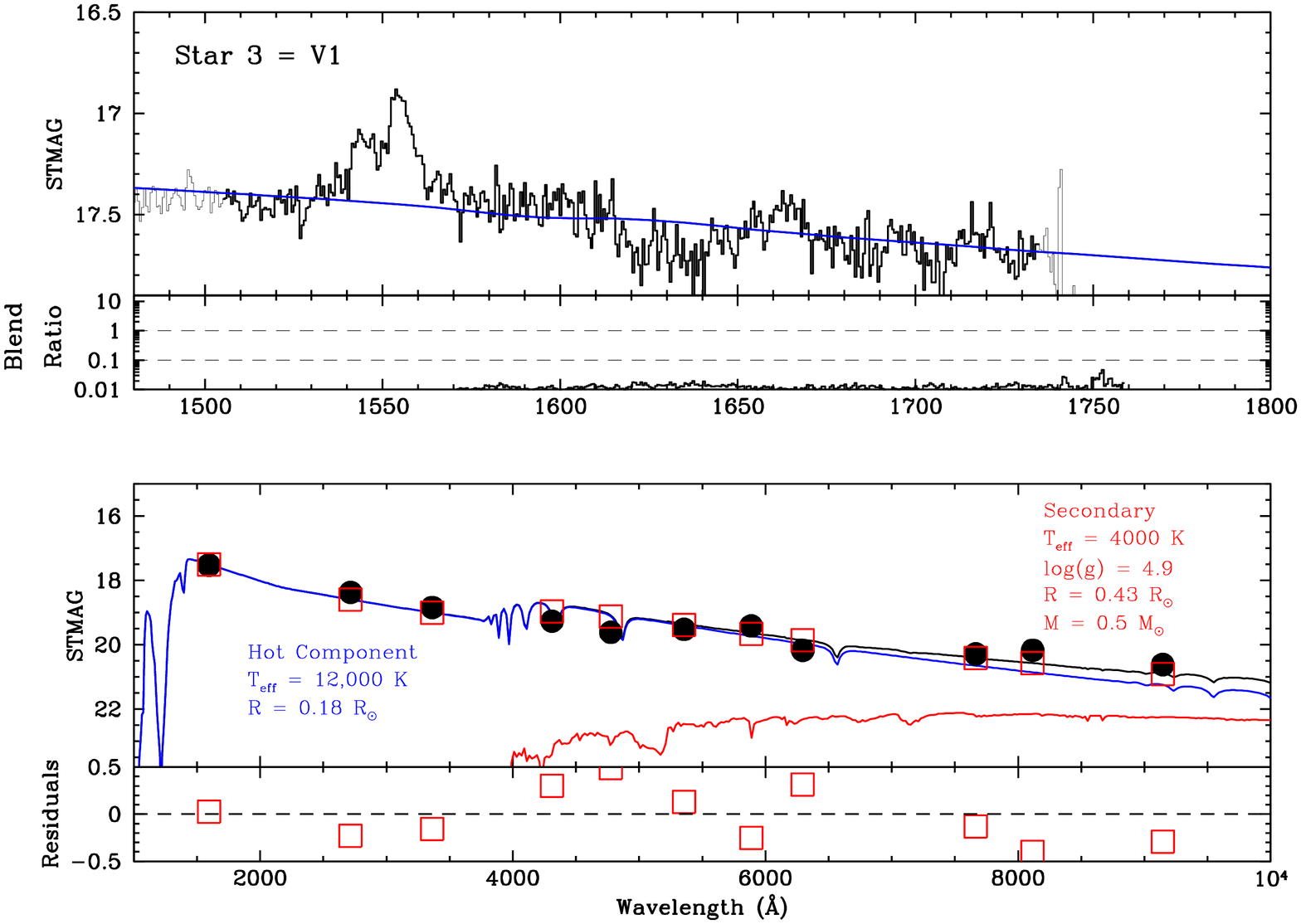}
\caption{The FUV spectrum (top panel) and broad-band SED (bottom
panel) of Star~3 = V1, along with our best-bet model fit to the
data (see text for details). Notation is as in 
Figure~\protect\ref{fig:star1}. Star~3 is a CV.}
\label{fig:star3}
\end{figure*}

Star~3 = V1 is a previously suspected CV
(Paresce, De Marchi \& Ferraro 1992; Grindlay et al. 2001a; Ferraro et
al. 2001; Knigge et al. 2002). Its orbital period is either 3.5 hrs or 7.0 hrs, 
depending on whether the observed optical variability is due to
ellipsoidal variations or not (Edmonds et al. 2003b). 

Figure~\ref{fig:star3} shows our FUV spectrum and broad-band SED for
V1. As already noted in Section~\ref{sec:gap_specs}, the detection of 
C~{\sc iv} in emission spectroscopically confirms V1 as a CV. The
emission line is double-peaked, which suggests that it was formed in a
rotating medium, such as an accretion disk (e.g. Smak 1981; Horne \&
Marsh 1986) or a rotating disk wind 
viewed at high inclination (e.g. Knigge, Woods \& Drew 1995). The FUV-NIR
SED is consistent with a single, hot component dominating the flux at
all wavelengths. 

Since the donor of V1 must fill its Roche lobe, we can use the
period-mean density relation (Equation~\ref{eq:rho}) to constrain its
nature. If the orbital period is 7.0~hrs, and the donor lies close to
the cluster MS, the donor mass would have to be roughly 0.75~$M_{\odot}$.
Such a star would be 0.5 mag - 1.0 mag brighter than
V1 at $\lambda \gtappeq 5000$~\AA; this makes the long-period option
unlikely. 
\footnote{Strictly speaking, unevolved CV donors are slightly bloated
relative to isolated MS stars (e.g. Knigge 2006). However, even if we
assume that the secondary in V1 is a full 20\% larger than a MS star 
(and adjust its parameters so as to still satisfy the
period-density relation), it would be significantly ($\simeq 0.3$~mag)
brighter than the observations at $\lambda \gtappeq 6000$~\AA. Any 
contribution from the hot component that must dominate at shorter
wavelengths would make this discrepancy even worse.} 
If the period is 3.5~hrs instead (which would imply that the
observed variability is not ellipsoidal in nature), the expected donor
mass is about 0.5~$M_{\odot}$. As shown in Figure~\ref{fig:star3},
such a donor would be a relatively minor contributor to the flux at
all FUV-NIR wavelengths. This is self-consistent, since a faint 
donor like this would not produce a significant ellipsoidal signal. We
therefore favour the 3.5~hr period and suggest that the donor lies on
or near the lower MS of the cluster. 

Both the FUV continuum and the FUV-NIR SED are reasonably described by
a 12,000~K optically thick component with an effective radius of
0.18~$R_{\odot}$ (where we have again used a $\log{\;g} = 7.0$\ WD model
to represent this component). Very similar models have been found to
match the accretion disk-dominated SEDs of nova-like CVs in the
Galactic field (e.g. Knigge et al. 1998). The identification of V1 as
a nova-like CV containing a bright accretion disk is consistent with
the fact that no dwarf nova outbursts have so far been discovered in
this source. However, we do note that the observed SED in
Figure~\ref{fig:star3} exhibits undulations (e.g. between 4000~\AA\ and
6000~\AA), which can also be seen as large (up to 0.5~mag) residuals
between model and data. These undulations and residuals are probably
due to variability. Even though nova-like CVs do not show the
eruptions associated with dwarf novae, they do produce a wide range of
variability, including flickering, orbital and even sizeable long-term
variations (e.g. Honeycutt, Robertson \& Turner 1998; Honeycutt \&
Kafka 2004; Knigge et al. 1998, 2000, 2004).

\subsection{Star~4: A Helium WD}

\label{sec:star4}

\begin{figure*} 
\center
\includegraphics[angle=-90,width=1.0\textwidth]{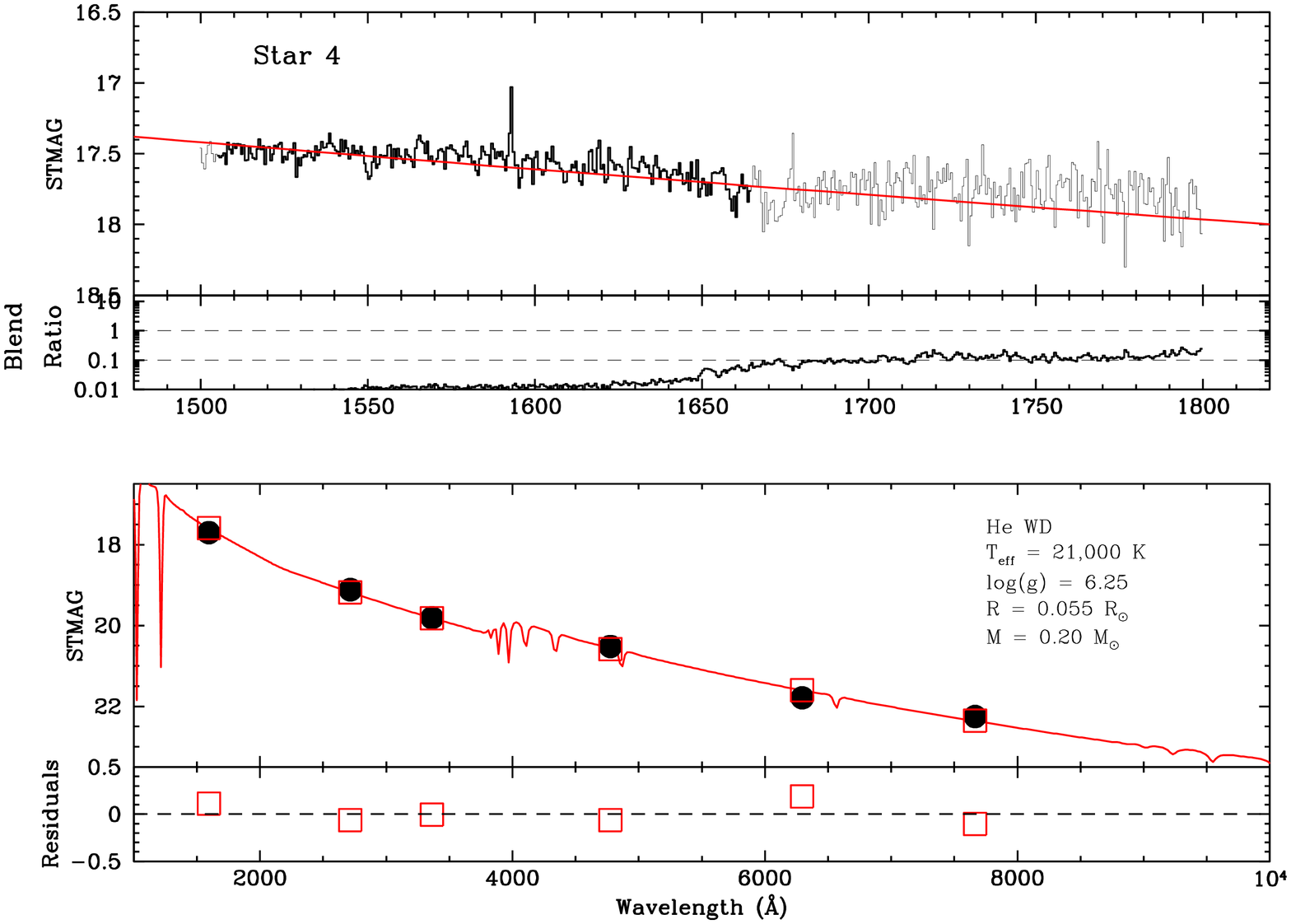}
\caption{The FUV spectrum (top panel) and broad-band SED (bottom
panel) of Star~4, along with our best-bet model fit to the
data (see text for details). Notation is as in 
Figure~\protect\ref{fig:star1}. Star~4 is a Helium white dwarf.}
\label{fig:star4}
\end{figure*}

Star~4 is a gap object photometrically, but one that lies close to the WD
cooling sequence in Figure~\ref{fig:cmd}. The FUV spectrum and
broad-band SED in Figure~\ref{fig:star4} are blue and featureless, but
it is impossible to fit these data with an ordinary WD model. The
observed spectral slope implies an effective temperature of around
20,000~K, but a normal WD with this temperature is much too faint to
match the observed flux. Instead, the required stellar radius is about 
0.05~$R_{\odot}$, 3 times larger than the radius of a CO-core WD. 

Such a radius is exactly what is expected for low-mass He-core WDs in
GCs (Serenelli et al. 2002). In fact, both the radius and 
temperature we infer for Star~4 are extremely similar to those of 
the He WD in NGC~6397 found by Edmonds et al. (1999). Motivated by
this, we fix the surface gravity of our WD model fit at $\log{\;g} =
6.25$ (Edmonds et al. 1999) and then find an excellent,
self-consistent description of the entire data set with $T_{eff} =
21,000$~K, $M = 0.2~M_{\odot}$ and $R_{\odot} = 0.055~R_{\odot}$. 

We conclude that Star~4 is a He WD in 47 Tuc. This is only the second
such object to be optically identified in this cluster, the other
being the low-mass companion to millisecond pulsar MSP-U (Edmonds et
al. 2001). Indeed, only a handful of He WDs have so far been optically
detected in all GCs combined, and the only ones not associated with
MSPs or ultra-compact X-ray binaries (e.g. Dieball et al. 2005b) are
the so-called  ``non-flickerers'' in NGC~6397 (Cool et al. 1998;
Edmonds et al. 1999) and the double-degenerate binary V46 in M4
(O'Toole et al. 2006). 

In order to form a He WD, its progenitor must lose sufficient mass to
avoid He ignition at the top of the RGB. The obvious way to truncate
the progenitor evolution is via mass loss during a mass-transfer
episode in a binary system. In a GC, He WDs can therefore be formed
either in primordial hard binaries or in binaries formed by dynamical
encounters (most likely exchange interactions). The channels available
for He WD formation in 47 Tuc have been discussed in detail by
Hansen, Kalogera \& Rasio (2003). They find that both the primordial
binary and exchange interaction channels can produce He WDs in 47
Tuc. In either case, the companion of the He WD is likely to be a NS
or CO-core WD. However, in a dense GC core, any He WD in a binary system
will be vulnerable to ejection during subsequent exchange
encounters. Thus isolated He WDs could be produced in GCs by dynamical
processes.

In the case of Star~4, there is certainly no sign of any binary
companion in the SED. However, only a radial velocity study will allow
us to determine if this is because the He WD is currently single or
because it has a dark, massive (NS or WD) companion. As noted by
Hansen, Kalogera \& Rasio (2003), He WDs are unique tracers of the
compact object populations in GCs and can provide insight into the
dynamical processes that drive GC evolution. Thus follow-up studies of
the known He WDs in GCs (including Star~4), as well as searches for
additional candidates, are likely to be extremely valuable.

%6752 -- Ferraro et al. 2003; Cocozza et al. 2006; Bassa et al. 2006 (MSP; association with cluster uncertain)
%M15 -- maybe Dieball....
%6397 -- Cool et al. 1998, Taylor et al. 2001; Edmonds et al. (1999)
%47 Tuc -- MSPs (e.g. Camilo et al. 2000) -- optically, MSP-U (Edmonds
%et al. 2001) -- Hansen, Kalogera & Rasio: could have He WDs formed by
%binary, exchange interactions involving NS and CO white dwarfs.
%M4 -- possibly V46 (O'Toole et al. 2005)

\subsection{Star~5: A Simple WD}
\label{sec:star5}

\begin{figure*} 
\center
\includegraphics[angle=-90,width=1.0\textwidth]{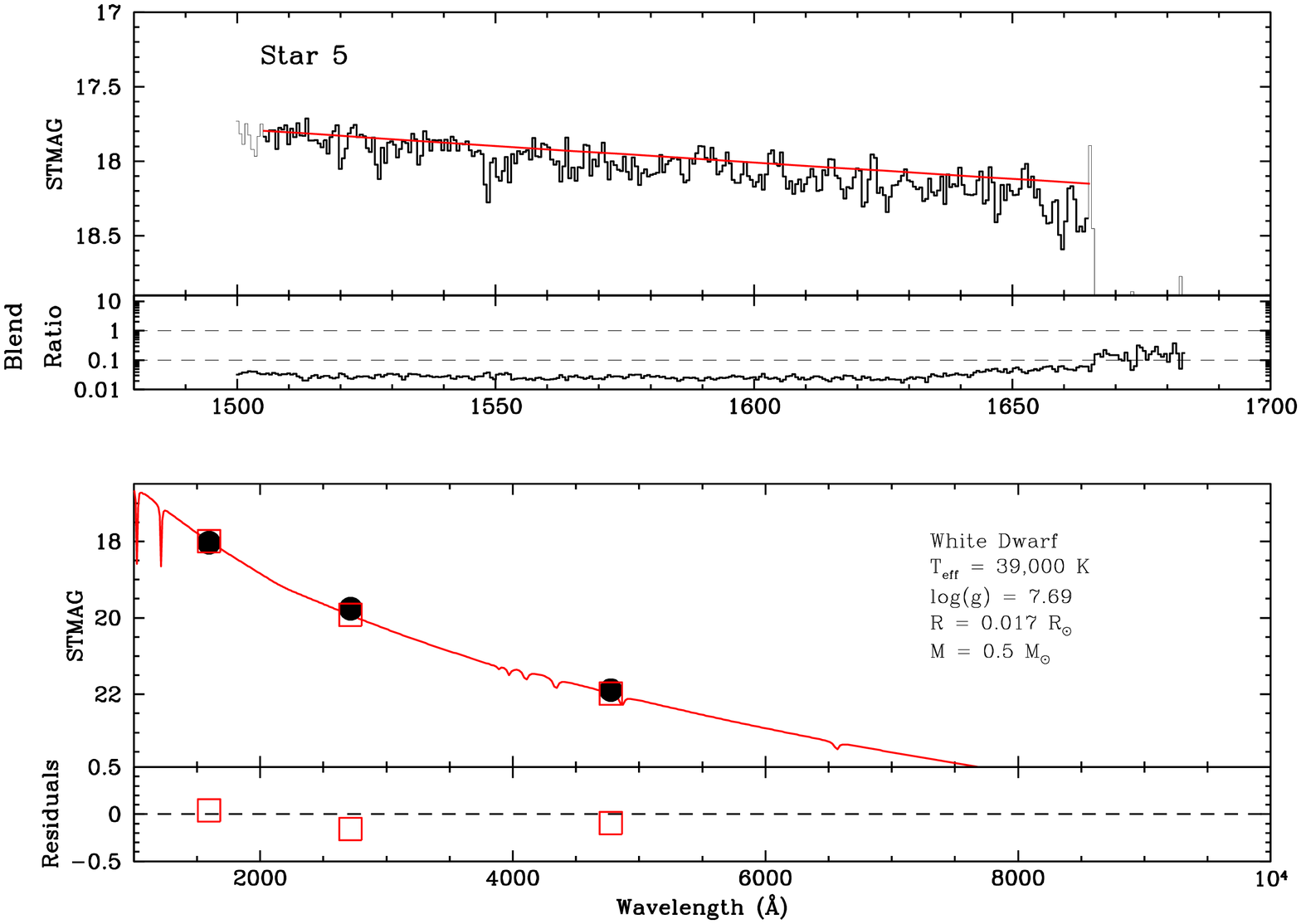}
\caption{The FUV spectrum (top panel) and broad-band SED (bottom
panel) of Star~5, along with our best-bet model fit to the
data (see text for details). Notation is as in 
Figure~\protect\ref{fig:star1}. Star~5 is a white dwarf.}
\label{fig:star5}
\end{figure*}

Star~5 is a representative example of hot, young WDs in 47~Tuc 
(c.f. Figure~\ref{fig:cmd}. The FUV spectrum and (very  
limited) broad-band photometry are shown in Figure~\ref{fig:star5}. The
object is too faint to be detected/measured in most of our optical
images, but all of the data are consistent with the SED expected for a
single, hot ($T_{eff} \simeq 39,000$~K) CO-core WD near the top of the
cooling sequence.

\subsection{Star~7: A WD-MS Binary}
\label{sec:star7}

\begin{figure*} 
\center
\includegraphics[angle=-90,width=1.0\textwidth]{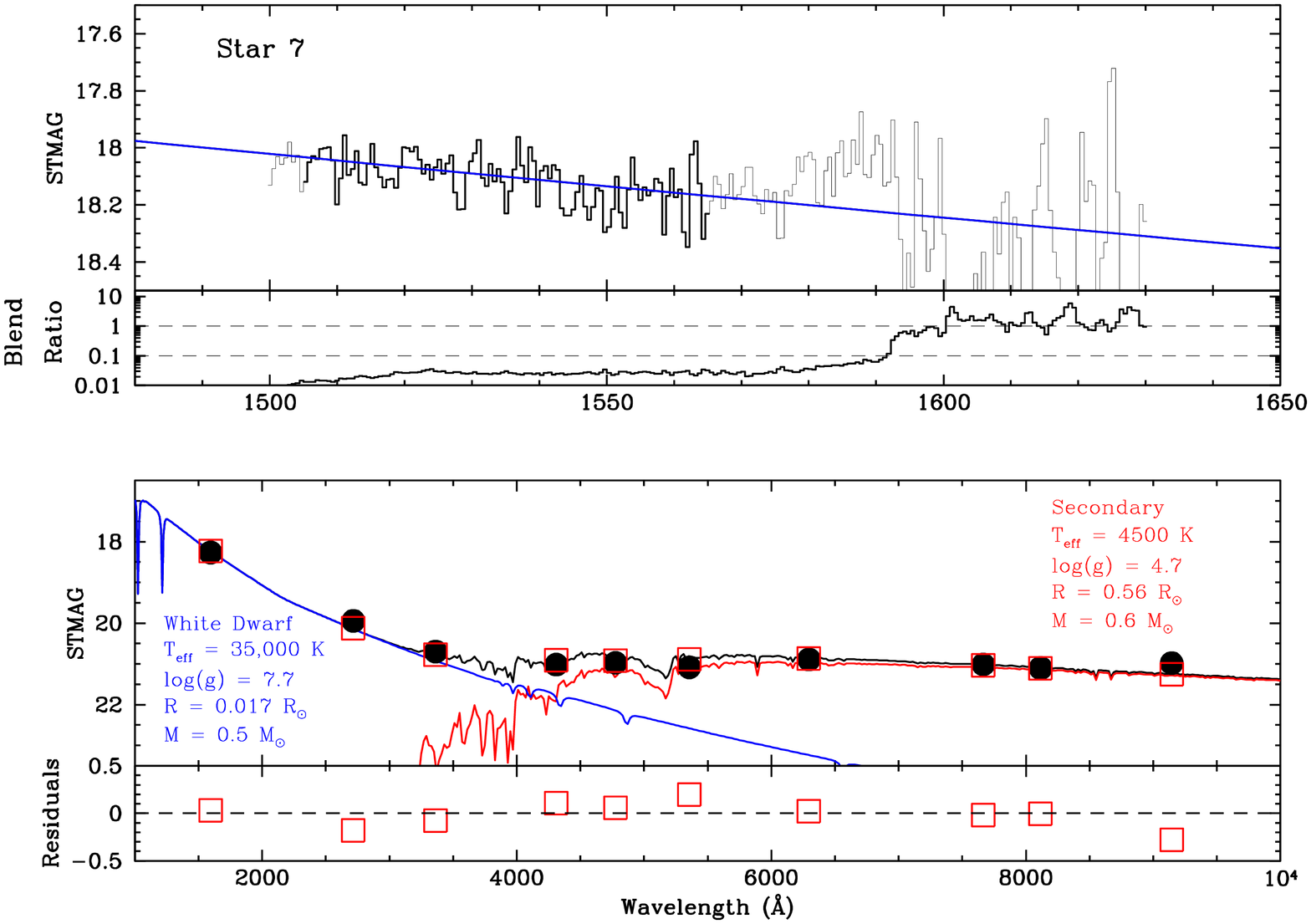}
\caption{The FUV spectrum (top panel) and broad-band SED (bottom
panel) of Star~7, along with our best-bet model fit to the data (see
text for details). Notation is as in
Figure~\protect\ref{fig:star1}. Star~7 is a WD-MS binary system.}  
\label{fig:star7}
\end{figure*}

Star~7 was originally included in this section as another
representative hot, young WD. However, despite its location close to
the WD cooling sequence in Figure~\ref{fig:cmd}, the broad-band
photometry in Figure~\ref{fig:star7} reveals a composite SED. As
expected from the CMD location, the WD dominates at short wavelengths
($\lambda \ltappeq 4000$~\AA). Beyond this, there is a red excess that
is well described by a 0.6~$M_{\odot}$\ MS star in the cluster. Even
though the expected number of false matches among FUV sources brighter
than $m_{FUV} = 19.5$\ is only $\simeq 0.3$\ (Section~\ref{sec:cmd}), it is,
of course, impossible to rule out completely that the composite SED
could arise from a chance coincidence between the WD and a low-mass MS
star. However, we see little evidence for blending at
intermediate wavelengths, where both components contribute
significantly to the total flux. We therefore believe Star~7 is 
likely to be a genuine WD-MS binary in 47~Tuc. 

\subsection{Star~10 = PC1-V36: A Close Binary with a Dark Primary and a Stripped Secondary}
\label{sec:star10}

\begin{figure*} 
\center
\includegraphics[angle=-90,width=1.0\textwidth]{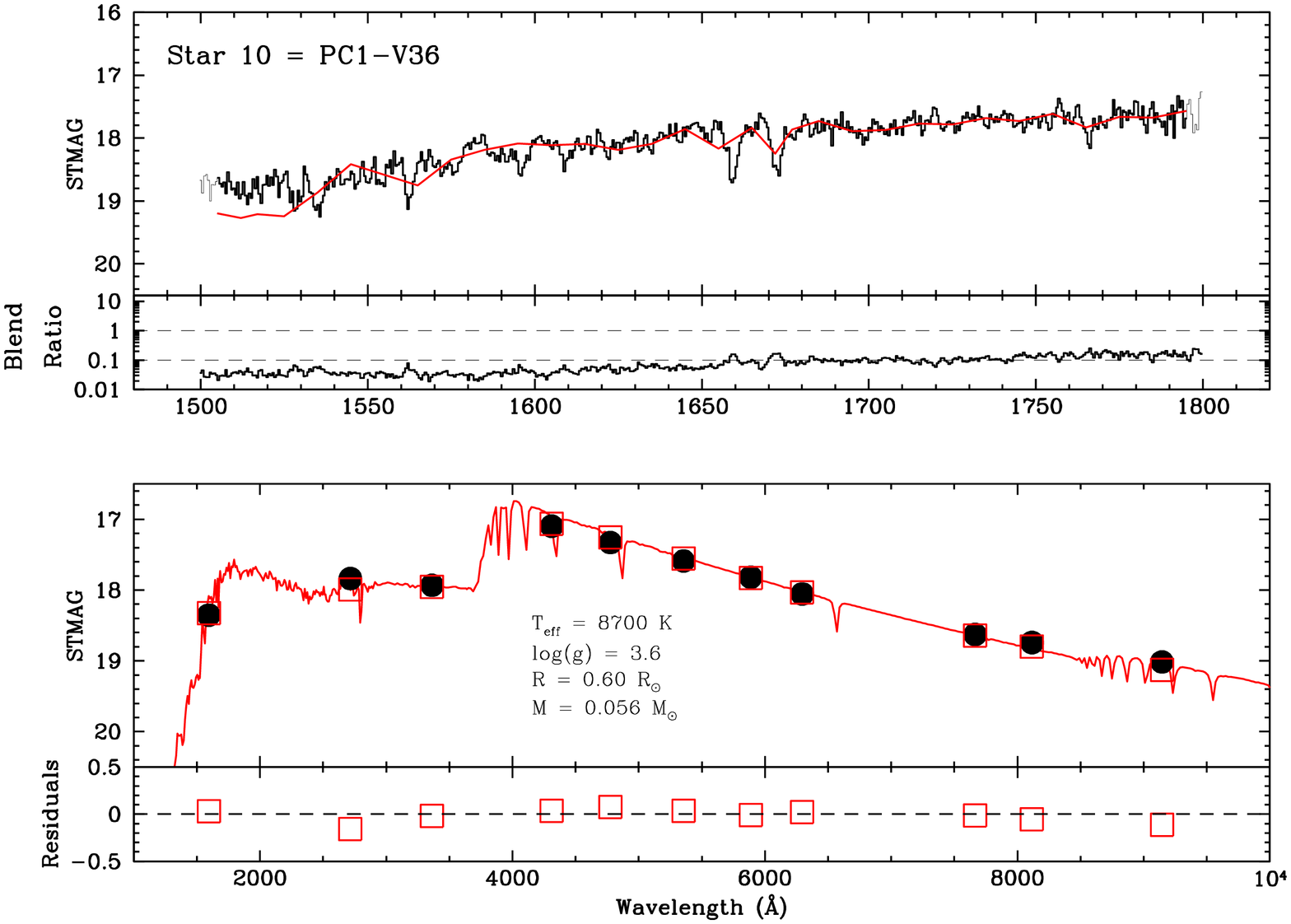}
\caption{The FUV spectrum (top panel) and broad-band SED (bottom
panel) of Star~10, along with our best-bet model fit to the
data (see text for details). Notation is as in 
Figure~\protect\ref{fig:star1}. Star~10 is a binary system whose SED
is dominated by a peculiar low-mass secondary star. This secondary is
probably the remnant of a subgiant that has been stripped of its
envelope.}  
\label{fig:star10}
\end{figure*}

Star~10 = PC1-V36 is arguably the most interesting and exotic object in our
entire sample. In Figure~\ref{fig:cmd}, it lies squarely in the gap
between the WD and main sequences, but we already noted in
Section~\ref{sec:gap_specs} that its FUV spectrum is rather
peculiar. In Figure~\ref{fig:star10}, we show this FUV spectrum again,
but now alongside with the broad-band SED we have constructed from the
ACS/HRC images. 

In one sense, the spectrum and FUV-NIR SED are extremely simple: a
single spectral component with $T_{eff} = 8700 \pm 40$~K, $\log{\;g} =
3.63 \pm 0.15$ and $R = 0.60\pm 0.01~R_{\odot}$ (corresponding to a
mass of $M = 0.056 \pm 0.018~M_{\odot}$) provides an excellent
description of our entire spectroscopic and photometric data
set. 
\footnote{We note in passing that an optical spectrum of Star~10 has
been presented and analyzed by De Marco et al. (2005). Their preferred
parameters for this object (N104-3 
in their notation) are $T_{eff} = 7650 \pm 250$~K, $\log{\;g} \simeq 4.2
\pm 0.3$, $R = 0.9 \pm 0.2~R_{\odot}$. Such a model is 
inconsistent with the data presented here. For example, a star with
these parameters would be more than 1.5~mag fainter in the FUV
waveband than we observe. This discrepancy could be a sign of
large-amplitude variability on long time scales. However, our own SED
shows little sign of this, despite being constructed from 
non-simultaneous observations (see
Figure~\protect\ref{fig:star10}). Another possibility is that the
spectrum used by De Marco et al. (2005) was affected by blending, which
could bias the inferred stellar parameters. While it would clearly be
important to distinguish between these alternatives, doing so is
beyond the scope of the present paper.} However, this combination of
parameters obviously does not describe any kind of normal star in
47~Tuc. This reflects the highly unusual SED of Star~10, which
does not appear to be composite, yet is clearly much bluer than that
of normal main sequence, red giant and horizontal branch stars in 47
Tuc, and much redder than that of hot WDs. Note that we provide errors
here, because the parameters of this object are so unusual and 
there is no evidence for systematic residuals that might seriously
bias the fit. Also, unlike for most other objects, no parameters were
constrained {\em a priori} in the fit.  

Perhaps the most obvious question raised by the strange SED is
whether Star~10 is actually a cluster member. Fortunately, this object 
is the only non-BS in our sample for which a reliable proper motion
measurement is available in McLaughlin et al. (2006). As shown in
Figure~\ref{fig:proper}, these data are entirely consistent with
membership of 47~Tuc.

Additional confirmation of the inferred parameters comes from the
variability of the object. Albrow et al. (2001) found a periodic
signal with $P = 0.4$~d in their optical time-series photometry for
Star~10 (=~PC1-V36 in their notation). The light curve shape was
consistent with ellipsoidal modulations, in 
which case the orbital period of the binary system would be $P_{orb} =
0.8$~d. If this object {\em is} an ellipsoidal variable, the star that
dominates the SED must be Roche lobe-filling and should obey the
orbital period-mean density relationship given in 
Equation~\ref{eq:rho}. Combining $P_{orb} = 0.8$~d with our
well-determined radius of $R = 0.60~R_{\odot}$, this relationship
predicts a mass of $M = 0.045~M_{\odot}$. This is completely
consistent with the mass inferred from the spectral/SED fit. 
We conclude that Star~10 is a 0.8~d binary system containing a 
bright, low-mass secondary star that fills (or nearly fills) its Roche
lobe. There is no sign of the primary in the SED, which rules out
MS-TO stars and  other optically bright objects.  

With an effective temperature higher than any BS, a mass below the
Hydrogen-burning limit, and a radius comparable to a MS star, the
secondary in the Star~10 binary system must be in an extreme and
short-lived evolutionary state. In our view, the most likely 
interpretation is that the secondary star is the remnant of a
subgiant whose Hydrogen envelope has been almost completely
stripped off. A similar description has been suggested for the
donor star in the X-ray binary AC~211 in M15 (van Zyl et al. 2004). 
The requisite stripping could have occurred as a result
of mass transfer in a primordial or dynamically-formed binary
system. Alternatively, and more excitingly, it might have occurred
during the very dynamical encounter that produced the binary
system (such as a physical collision with a NS; e.g. Lombardi et
al. 2006). In this case, Star~10 might be the ``smoking gun'' of a 
recent dynamical encounter involving a compact object.

\subsection{Star~15: A WD-Subgiant Binary?}
\label{sec:star15}

\begin{figure*} 
\center
\includegraphics[angle=-90,width=1.0\textwidth]{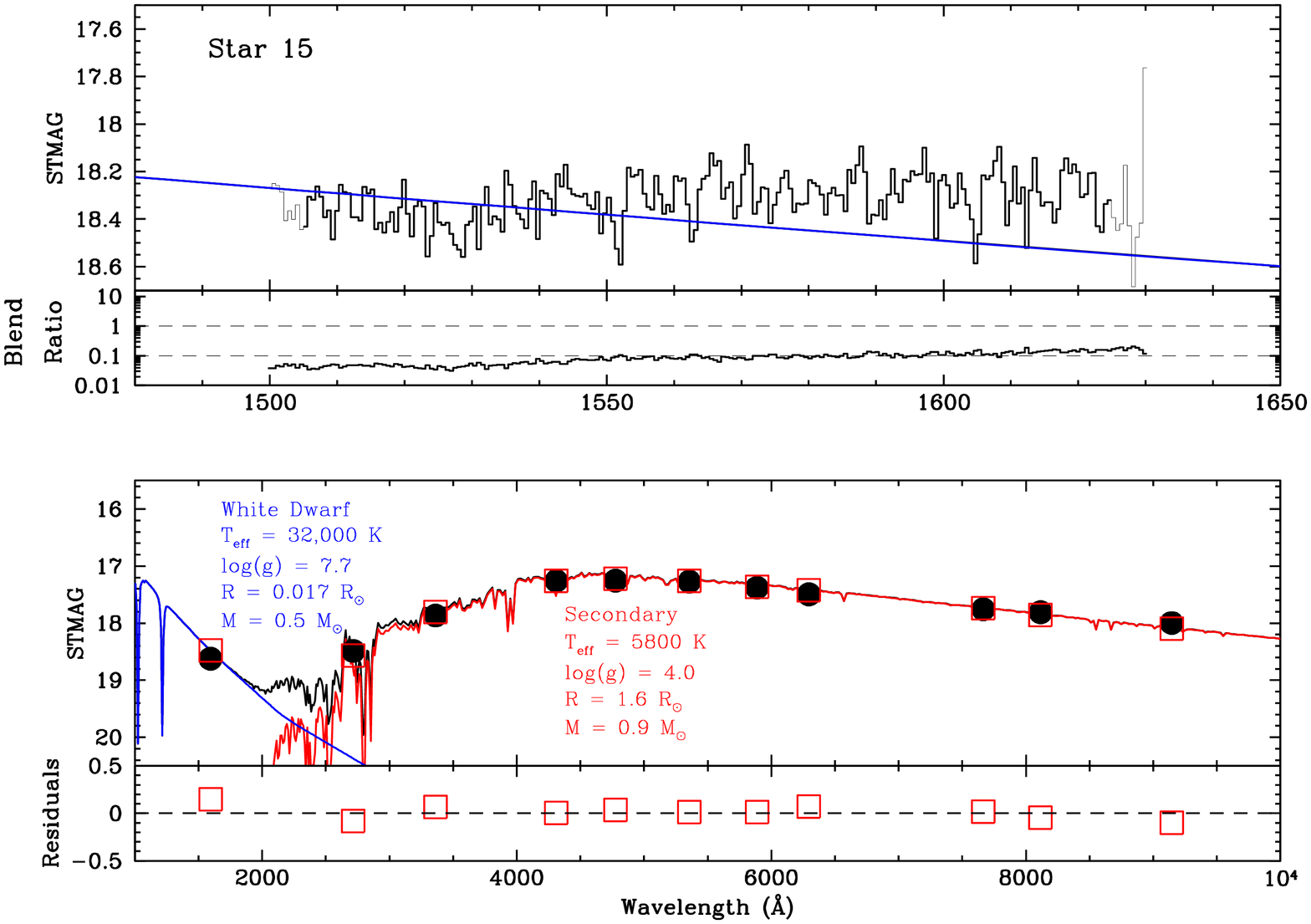}
\caption{The FUV spectrum (top panel) and broad-band SED (bottom
panel) of Star~15, along with our best-bet model fit to the data (see
text for details). Notation is as in
Figure~\protect\ref{fig:star1}. Star~15 is either a WD-subgiant binary
system or a chance superposition of two such stars.}
\label{fig:star15}
\end{figure*}

Star~15 is a gap source and located very close to Star~10~=~PC1-V36 in
the FUV/optical CMD (Figure~\ref{fig:cmd}). However, its FUV spectrum and
broad-band SED in Figure~\ref{fig:star15} reveal that its properties
are completely different from those of Star~10. Our suggested model
for this system consists of a hot WD (which dominates the FUV) and a
MS-TO/subgiant secondary star (which dominates everything else). This
describes 
the data fairly well, although the observed FUV spectrum is somewhat
flatter than expected for a WD that matches the FUV flux level. We
have not attempted to fit any more complicated models, since the FUV
spectrum suffers from at least mild blending across much of the
spectral range.  

As usual, we have to allow for the possibility that the apparently
composite nature of Star~15 is due to a chance coincidence. Even
though the expected number of false matches among FUV sources brighter
than $m_{FUV} = 19.5$\ is only $\simeq 0.3$\ (Section~\ref{sec:cmd}), 
this concern deserves particular attention in this case, because the offset
between FUV and optical positions listed in Table~\ref{tab:specdat} is
relatively large (1.29 FUV pixels). While this is well within the
hard cut-off adopted in Paper~I (1.5~pixels) that was also used for
estimating the expected number of chance coincidences, the probability
of a {\em given} match being due to chance clearly increases with
increasing offset. Based on careful visual inspection and  
blinking of the FUV and optical images, we think there is a strong 
possibility that the FUV and optical sources associated with Star~15 
will turn out to be unrelated, altough it is impossible to be
certain. In order to flag this, we have 
therefore added a ``?''  to the spectroscopic/SED classification in
Table~\ref{tab:specdat}.

\subsection{Star~17: A Probable WD-Subgiant Binary and Possible CV}
\label{sec:star17}

\begin{figure*} 
\center
\includegraphics[angle=-90,width=1.0\textwidth]{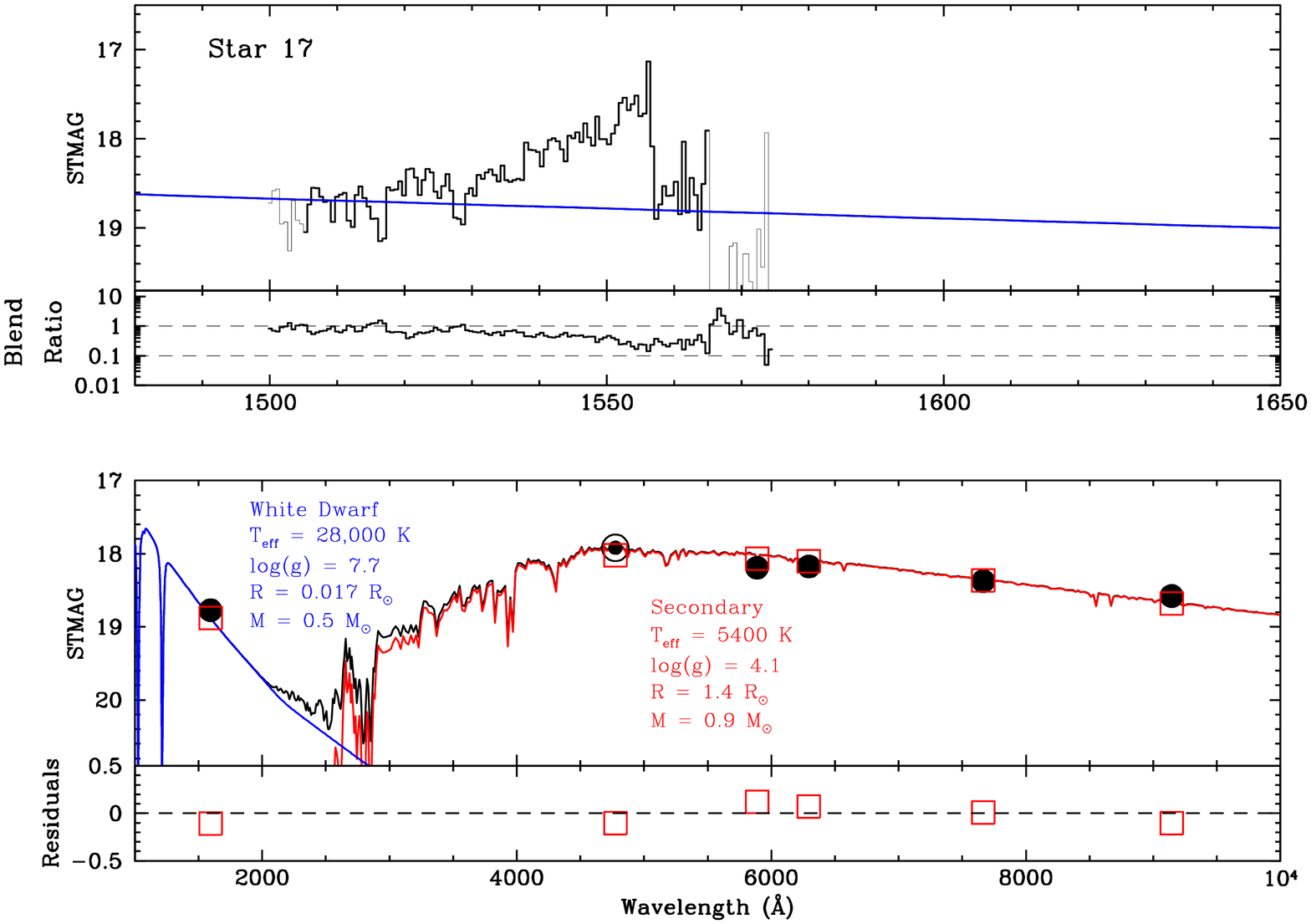}
\caption{The FUV spectrum (top panel) and broad-band SED (bottom
panel) of Star~17, along with our best-bet model fit to the data (see
text for details). The data point near 4800~\AA\ was taken from
McLaughlin et al. (2006) and is therefore shown with a different
symbol. Other notation is as in
Figure~\protect\ref{fig:star1}. Star~17 is a probable WD-subgiant
binary system  and possible CV (based on the tentative detection of
C~{\sc iv} in emission).}
\label{fig:star17}
\end{figure*}

Star~17 is another gap source and located close to Stars~10 and 15 in
the FUV/optical CMD (Figure~\ref{fig:cmd}). Its broad-band SED
(Figure 18, bottom panel) is also
very similar to that of Star~15 and suggests that Star~17 is another
WD-subgiant binary system.  However, as already noted in
Section~\ref{sec:gap_specs}, there is marginal evidence that C~{\sc
  iv} 1550~\AA\ is in emission in this sytem (Figure 18, top panel). The
detection is not 
compelling, partly because the line happens to lie close to the edge
of the detector and is therefore cut off, and partly because the
spectrum of Star~17 suffers from moderate blending. 

In assessing the likelihood that Star~17 may be a previously unknown
CV in 47~Tuc, it is worth noting that this FUV source was not found
to be variable in Paper~I, and is also not a viable counterpart to any
of the Chandra X-ray sources in Heinke et al. (2005). On balance, we
are currently skeptical that this system will be confirmed as
a CV in the future. 

The status of Star~17 as a WD-subgiant binary is somewhat more secure,
although here again we have to consider the possibility that the match
between FUV and optical sources is due to a chance coincidence. We
note again that the expected number of false matches among FUV sources
brighter than $m_{FUV} = 19.5$\ is only $\simeq 0.3$\ 
(Section~\ref{sec:cmd}). However, as in the case of Star~15, the
offset between FUV and optical positions is relatively large for
Star~17 (1.20~FUV pixels), so the possibility of a false match
deserves to be taken seriously nonetheless. We have therefore again
inspected and blinked the FUV and optical images in the vicinity of
Star~17. While we cannot rule out the possibility of a random match
based on this, we do think Star~17 is more likely to be a genuine
match than Star~15. Statistically, it would certainly be rather
surprising if {\em both} Stars~15 and 17 turned out to be chance
coincidences. Nevertheless, we have again conservatively marked the
binary classification for Star~17 in Table~\ref{tab:specdat} with a
``?''.

\subsection{Star~20~=~V2: A CV with a Main-Sequence Donor}
\label{sec:star20}

\begin{figure*} 
\center
\includegraphics[angle=-90,width=1.0\textwidth]{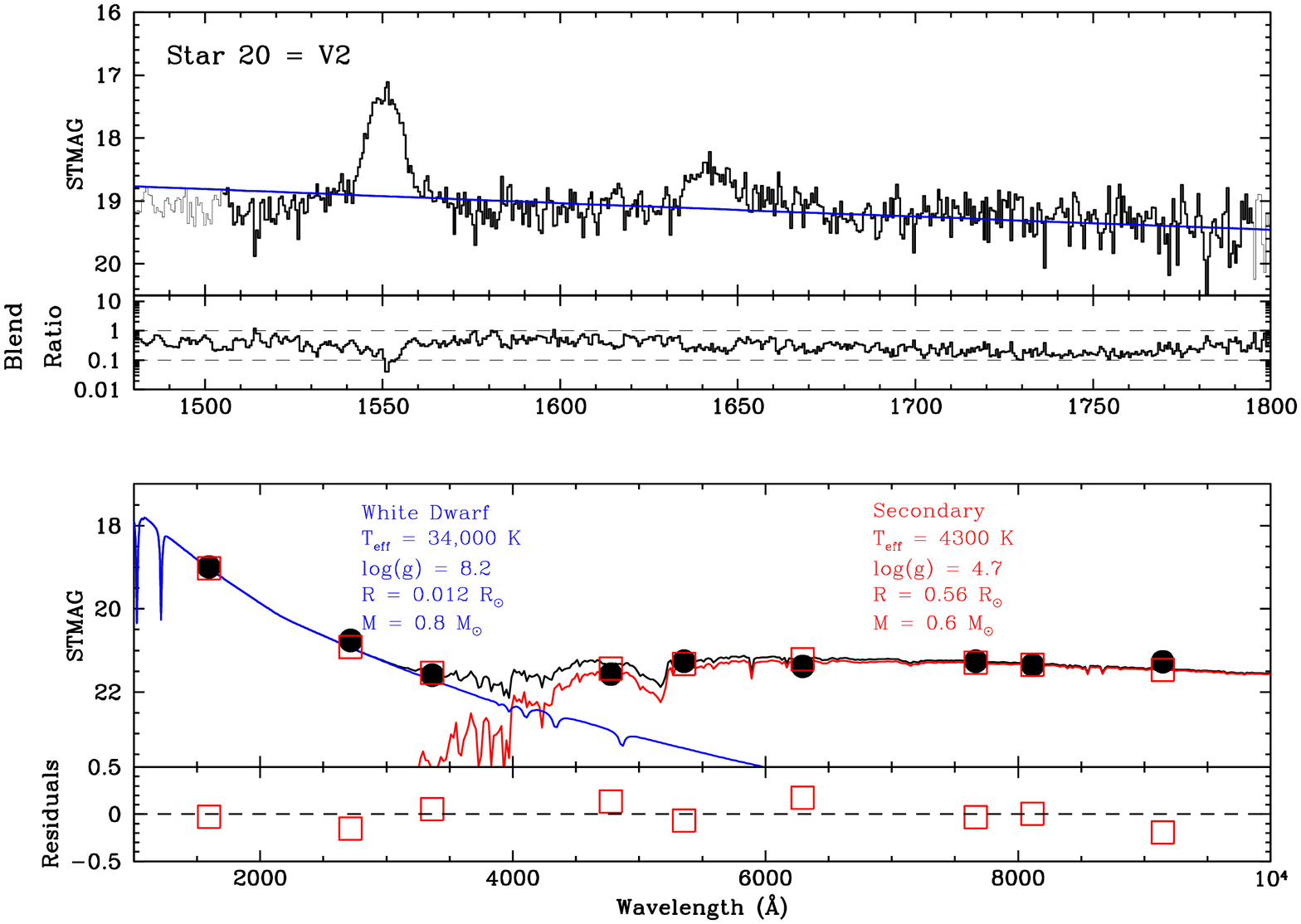}
\caption{The FUV spectrum (top panel) and broad-band SED (bottom
panel) of Star~20, along with our best-bet model fit to the data (see
text for details). Notation is as in
Figure~\protect\ref{fig:star1}. Star~20 is a CV.}
\label{fig:star20}
\end{figure*}

Star 20~=~V2 is a previously known CV that was originally discovered as an
erupting dwarf nova by Paresce \& De Marchi (1994). A second outburst
was found by Shara et al. (1996). V2 is also known to be an X-ray source
(Grindlay et al. 2001a), as well as a UV-excess object located in the
gap region of the CMD (Ferraro et al. 2001; Knigge et
al. 2002). 

The FUV spectrum and broad-band SED we have obtained for V2 are shown
in Figure~\ref{fig:star20}. As already noted in
Section~\ref{sec:gap_specs}, the presence of C~{\sc iv}~1550~\AA\ and
He~{\sc ii}~1640~\AA\ emission lines spectroscopically confirms the CV
classification of this source. The SED can be decomposed into a hot
component with parameters appropriate for a massive WD, and a cool
component with parameters appropriate for a low-mass MS star in 47
Tuc. 

Edmonds et al. (2003b) found evidence for variability on periods of 
6.8~hrs and 3.0~hrs in their optical photometry for V2. Allowing for
the possibility that one of the signals is ellipsoidal in nature (in
which case the orbital period is twice the observed period), this
means that there are four candidate orbital periods for this
system. For comparison, the donor parameters suggested by our SED fit
(in which we constrained the donor to lie on the cluster MS), would
imply an orbital period of 4.8~hrs (via Equation~\ref{eq:rho}). We
will not pursue this issue further here, since neither the orbital
period nor our estimate of the donor properties is secure. We do note
that donors with significantly higher mass than that implied by our
fit are excluded, since they would be brighter than the observed SED
at optical and NIR wavelengths. Thus the broader classification of the
secondary as a lower MS star {\em is} fairly secure. 

\begin{figure*} 
\center
\includegraphics[angle=0,width=0.95\textwidth]{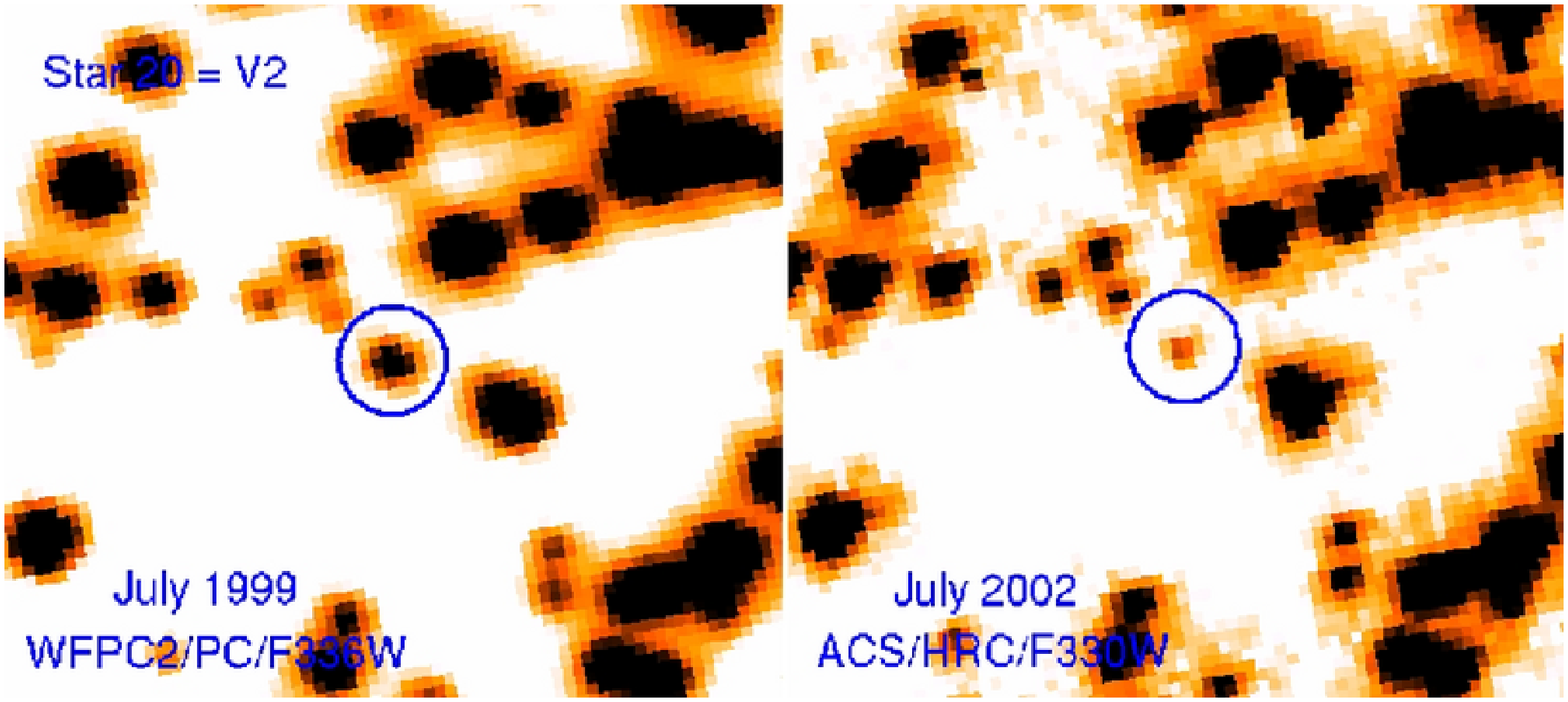}
\includegraphics[angle=0,width=0.95\textwidth]{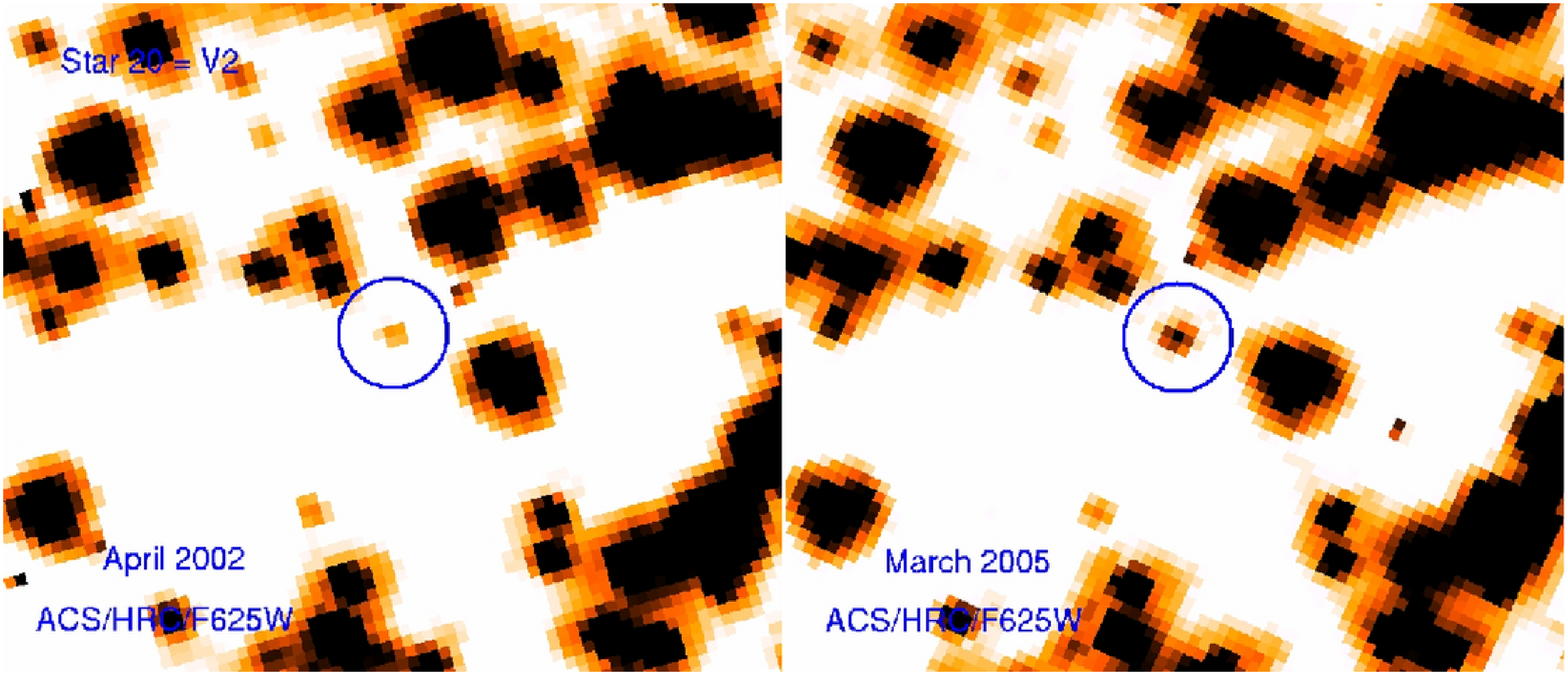}
\caption{
{\em Top Row:} Approximately U-band observations of
Star~20~=~V2 obtained with {\em HST} in July 1999 (left panel) and July 2002
(right panel). The difference in brightness is obvious and implies
that V2 experienced a dwarf nova eruption in or near July 1999.
{\em Bottom Row:} Approximately R-band observations of
Star~20~=~V2 obtained with {\em HST} in April 2002 (left panel) and March
2005 (right panel). The difference in brightness is obvious and implies
that V2 experienced a dwarf nova eruption in or near March 2005. In all panels, 
north is up, east is to the left, and the field of view is approximately 
1.8\arcsec.}   
\label{fig:v2_outburst}
\end{figure*}

In the process of constructing our broad-band SED, we also found
evidence of two additional eruptions of V2, which are shown in
Figure~\ref{fig:v2_outburst}. First, the ACS/HRC/F330W data
point in Figure~\ref{fig:star20} is 1.7~mag fainter than the older
WFPC2/PC/F336W measurement that was used in the construction of the
CMD (Figure~\ref{fig:cmd}. This suggests that V2 was in outburst in
July 1999, the epoch that dominates in our co-added PC/F336W
image. This is consistent with the fact that Ferraro et al. (2001)
report an even brighter WFPC2/PC/F336W magnitude for V2 from data
obtained exclusively in July 1999. The fainter ACS/HRC/F330W
measurement would move V2 much closer to the WD sequence in
Figure~\ref{fig:cmd}. Second, there is a 1.0 mag difference between
the two ACS/HRC/F625 magnitude estimates we have for this source. The
bright estimate comes from an image taken in March 2005, which is the
only image obtained at this time in our optical data set. The
discovery of two eruptions in our limited optical data set confirms
that V2 is a dwarf nova with a relatively high duty cycle (c.f. Shara
et al. 1996).

\subsection{Star~27: An SMC Interloper}
\label{sec:star27}

\begin{figure*} 
\center
\includegraphics[angle=-90,width=1.0\textwidth]{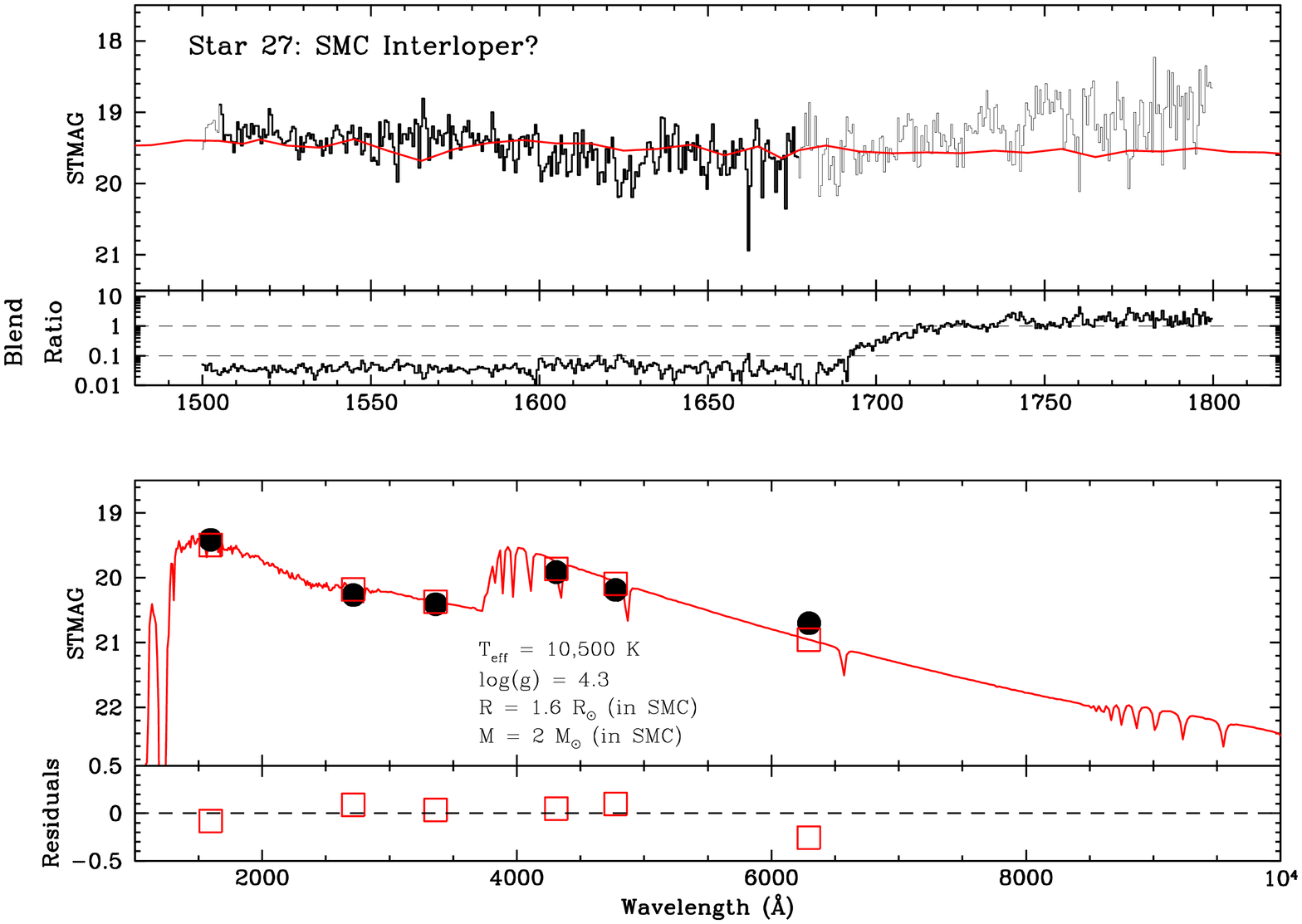}
\caption{The FUV spectrum (top panel) and broad-band SED (bottom
panel) of Star~27, along with our best-bet model fit to the data (see
text for details). Notation is as in
Figure~\protect\ref{fig:star1}. Star~27 is a massive star that is
located behind 47~Tuc in the SMC}
\label{fig:star27}
\end{figure*}

Star~27 is another gap source in Figure~\ref{fig:cmd}, and its FUV
spectrum and broad-band SED are shown in
Figure~\ref{fig:star27}. Despite several attempts, we were unable to
find a physically plausible one- or two-component fit to this data
that was consistent with the source being a cluster member. 

We therefore considered the possibility that Star~27 is a background
star. As discussed in Paper~I, the outskirts of the SMC are located
directly behind 47~Tuc, and roughly one SMC interloper may be expected to
contaminate our FUV/optical CMD. Figure~\ref{fig:cmd} shows that
Star~27 is a good candidate, since it lies close to the expected
location of the SMC main sequence. Note that this object is not
included in Figure~\ref{fig:proper}, since there is no proper motion
information for it in McLaughlin et al. (2006). Thus a location in the
SMC cannot be excluded for Star~27.

In order to test this idea, we changed the distance adopted in
our fits to 60~kpc (the average metallicity in the SMC is similar to
that of 47~Tuc). We then find that a single-component model with $T_{eff}
\simeq 10,500$~K, $R = 1.6~R_{\odot}$ and $\log{\;g} = 4.3$ provides a
good description of all of the datal. These parameters are entirely
reasonable for a slightly evolved $2 M_{\odot}$\ star in the SMC. We
therefore conclude that Star~27 is an SMC interloper and unrelated to
47~Tuc. 

\subsection{Star~999: A Bright and Probably Massive Blue Straggler}
\label{sec:star999}

\begin{figure*} 
\center
\includegraphics[angle=-90,width=1.0\textwidth]{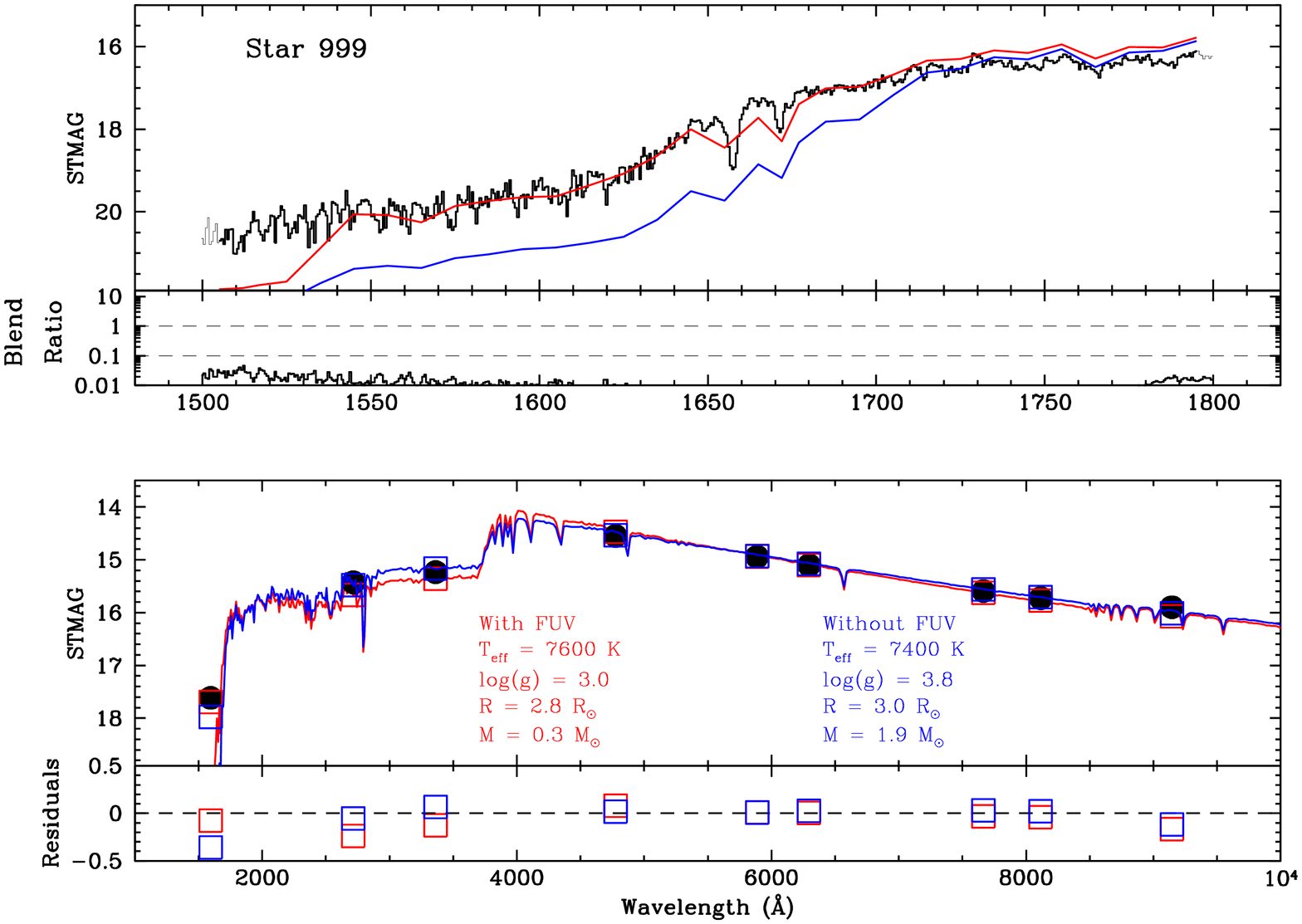}
\caption{The FUV spectrum (top panel) and broad-band SED (bottom
panel) of Star~20, along with two model fits to the data (see
text for details). Notation is as in
Figure~\protect\ref{fig:star1}. Star~999 is a blue straggler.} 
\label{fig:star999}
\end{figure*}

Star~999 is the optically brightest BS in our spectroscopic
sample. Its FUV spectrum and broad-band SED are shown in
Figure~\ref{fig:star999}, and imply $T_{eff} \simeq 7500$~K and $R
\simeq 3~R_{\odot}$\ for this object. These are reasonable parameters
for a BS in 47~Tuc. However, in order to obtain an acceptable single
component fit to all of the data, a surprisingly low surface gravity
is also required ($\log{\;g} \simeq 3.0$; red line in
Figure~\ref{fig:star999}. Taken at face value, this combination of
radius and surface gravity would point to an extremely low mass of $M
\simeq 0.3~M_{\odot}$\ for Star~999. This does not seem reasonable for
a BS.

There could be physical explanations for such an abnormally low
surface gravity, such as fast rotation (Porter \& Townsend 2005;
Knigge et al. 2006) or the presence of a disk around the star (De
Marco et al. 2004). However, before resorting to such explanations, it
is worth checking that the spectral models we are using can actually
be trusted in the regime we are using them here. The obvious worry
here is the reliability of the models in FUV region, since this
waveband is on exponential Wien tail of the BS SED where even
relatively slight inaccuracies in the model may result in large flux
changes.  

The work by Castelli \& Cacciari (2001) suggests that this worry is
well founded. They carried out an extensive FUV analysis of Pop~II
A-stars, based on the same set of Kurucz models that we employ 
here. Their main finding was that, for $T_{eff} \ltappeq 8700$~K,
model fits to the FUV spectra produce biased surface gravity estimates 
that are systematically lower than those implied by optical
data. For hotter stars, there was no systematic discrepancy between
the two sets of estimates.

As a test, we therefore repeated the one-component fit, but excluded
the FUV spectrum and photometry. The result of this exercise 
confirmed our suspicions: the best model fit to the NUV-NIR SED (the
blue line in Figure~\ref{fig:star999}) has a similar temperature and
radius to the global best fit, but a much higher surface gravity
($\log{\;g} = 3.8$). The BS mass implied by this fit is therefore also
much higher ($M \simeq 1.9~M_{\odot}$). 

Two points regarding the fits carried out in previous sections are
worth making at this stage. First, the effective  temperature of
Star~10 is just about high enough for our model fit to 
yield a meaningful surface gravity estimate. Second, the FUV excess seen
in the BS Star~2 (and attributed to a WD companion) cannot be
attributed to a similar model failure. This can be seen by comparing
the FUV properties of Stars~2 and 999 directly. The broad-band FUV
magnitude of both stars is dominated by a BS with $T_{eff} \simeq
7500$~K, with Star~2 being only 0.2~mag brighter than
Star~999. However, at the shortest wavelengths available in the FUV
spectra (1500\AA\ - 1600\AA), 
Star~2 is a full 2~mag brighter than Star~999. Thus the massive FUV
excess exhibited by Star~2 is not just relative to theoretical models,
but also relative to other BSs in the cluster. 

Returning to Star~999 itself, it is interesting to note that the mass
inferred from our fit to the NUV-NIR SED is just over twice the
turn-off mass in 47~Tuc ($M_{TO} \simeq 0.9~M_{\odot}$). This is 
interesting, since the formation of a BS with $M > 2 M_{TO}$\ must
involve three progenitors. Such a 
``supermassive'' BS (SMBS) cannot be formed in a single collision
between two MS stars, nor via mass transfer in an ordinary binary
system. Unfortunately, the uncertainties on the mass of Star~999 are
too large to establish it as a SMBS (the statistical error alone is 
about 0.4~$M_{\odot}$). The sample of BSs analysed by De Marco et
al. (2005) also contained several candidate SMBSs (including one in
NGC~6793 for which Saffer et al. [2002] had derived a preliminary mass
estimate of about $3~M_{TO}$). However, none of these candidates could
be confirmed conclusively as SMBSs. At the moment, this leaves the
BS-MS binary discovered by Knigge et al. (2006) as the 
most convincing candidate for a BS system with three progenitors in a GC
core. However, the first definitive SMBS remains to be discovered. 

\section{Discussion}
\label{sec:discuss}

Since we have discovered and analysed a fair number of interesting and
exotic objects in 47~Tuc, it is worth considering some of the wider 
implications of our results. In the following sections, we will therefore
take a look at three key issues: (i) the binary fraction in 47~Tuc;
(ii) the nature and abundance of CVs in GCs; (iii) the significance of
systems that are unique even by the standard of most other exotica
(such as Star~10 in our sample).

\subsection{The White Dwarf Binary Fraction in 47~Tuc}
\label{sec:binfraction}

We can use our results to obtain a rough estimate of the binary
fraction amongst hot, young WDs in 47~Tuc. As a starting point, we
note that our FUV photometry 
includes 25 objects brighter than $m_{FUV} = 19.5$\ and within our
spectroscopic field of view. Figure~\ref{fig:cmd} shows that, above
this limit, we have useable spectra for all gap objects, all but one
BSs and all objects on the WD cooling sequence. Thus we use $m_{FUV} =
19.5$\ as an approximate completeness limit for the discovery of WD
binary systems in our spectroscopic sample. 

There are 5 objects without optical counterparts above this limit, as
well as 5 objects on the WD cooling sequence for which we have no
evidence of a companion. As a check, we visually inspected the
vicinity of each of these 10 FUV sources in the deepest red optical 
image available. For most objects, this was a 600~s F814W exposure. In
a few cases, the location of interest fell outside the F814W field of
view or on an image defect, so we used a 
350~s F555W exposure instead.\footnote{One object -- Star~6 -- was
outside the field of view of both images, so in this case we
inspected a shallower 60~s F814W exposure.} No additional red
counterparts were found, down to an estimated limit of $m_{STMAG}
\simeq 21$\ or better. This excludes the existence of MS companions
as bright as that in Star~7.

Thus there are probably 10 single WDs in our bright ($m_{FUV} <
19.5$) sub-sample. This same sub-sample also contains two very
probable binary systems containing hot, young WDs 
(Stars~2 and 7), and one additional possible one (Star~15). CVs have
been excluded here, because their WDs are probably 
accretion-heated, rather than young. We also exclude Star~17, partly
because it is a candidate CV, and partly because it, too, could be a
chance superposition. A rough estimate 
of the binary fraction amongst hot WDs in the core of 47~Tuc is then
2/13 (assuming that Star~15 is {\em not} a binary), i.e. $f_{bin}=
15\%~^{+17\%}_{-9\%}~{\rm (stat)}~^{+8\%}_{-7\%}~{\rm (sys)}$. The
statistical error on this estimate is solely due to small 
number statistics, whereas the quoted systematic error shows how the
maximum-likelihood estimate would change if the true number of WD
binaries in our sample was 1 or 3, instead of 2.

Clearly, this estimate needs to be viewed with great caution. Aside
from the issues of small number statistics and chance coincidences, we 
cannot rule out the possibility of faint, lower MS companions to some
of the apparently single WDs in our sample. On the other hand, the
number of WDs (objects near the 
cooling sequence in Figure~\ref{fig:cmd}, as well as objects without a 
F336W counterpart) rises considerably more steeply towards fainter
magnitudes than the numbers of gap sources and BSs. Since Stars~2 and
15 were found amongst the latter two classes, our binary candidate
haul may have been unusually lucky.  

While the allowed range we have determined is not yet
very constraining, it is certainly consistent with the much more accurate
estimates of the binary fraction determined by 
Albrow et al. (2001) in the core of 47~Tuc. They found $13\% \pm6\%
$\ based on the number of eclipsing binaries in their data, and
$14\%\pm4\% $\ based on the number of W~UMa  
stars. Even though the estimate presented here is much less accurate,
it is the first to be derived from the WD  population of a GC and
demonstrates the 
potential of FUV surveys in this area. Given the importance of core
binary fractions to cluster dynamics -- and the present controversy
surrounding the theoretical predictions (Ivanova et al. 2005; Hurley,
Aarseth \& Shara 2007; Fregeau 2007) -- more accurate WD-based
estimates could be extremely valuable. 

\subsection{Cataclysmic Variables in 47 Tuc and other GCs}
\label{sec:CVs}

Theoretical models for CV formation in GCs predict that a cluster like
47~Tuc should harbour $\sim 200$, mostly dynamically-formed CVs (di
Stefano \&  Rappaport 1994; Ivanova et al. 2006; but also see Townsley
\& Bildsten 2005). Even though 
recent surveys have at last begun to uncover a sizeable population of
CV candidates in GCs, there is still a clear shortfall in the observed
numbers relative to these predictions. For example, the combination
of deep Chandra X-ray surveys and optical follow-up has so far led
to the identification of $\sim 20$~probable CVs in 47~Tuc (Grindlay et
al. 2001a; Edmonds et al. 2003ab; Heinke et al. 2005). However, it is
still not clear if this discrepancy points to a problem with our
understanding of CV formation and/or evolution, or if it is simply due 
to observational incompleteness (see Heinke et al. 2005, Ivanova et
al. 2006 and Maccarone \& Knigge 2007 for discussions of this point). 

The present results are relevant to this issue for several
reasons. First, the discovery of emission lines in three previously
suggested CV candidates in our spectroscopic sample 
(Star~1~=~AKO~9; Star~3~=~V1; Star~20~=~V2) confirms that combined X-ray-,
colour- and variability-based searches are, in fact, finding genuine
CVs. This is rather important, since only seven other CVs have so far
been spectroscopically confirmed in all Galactic GCs combined.
\footnote{These are CVs 1-4 in NGC~6397 (Grindlay et al. 1995; Edmonds
  et al. 1999), V101 in M5 (Margon, Downes \& Gunn 1981; Naylor et
  al. 1989; Shara, Moffat \& Potter 1990); Nova 1938 in M4 (Shara,
  Moffat \& Potter 1990); and Star~1 in NGC~6624 (Deutsch et
  al. 1999). Note that the spectroscopically confirmed dwarf nova V4
  in the field of M30 (Margon \& Downes 1983; Shara, Moffat \& Potter
  1990) is probably a foreground object (Machin et al. 1991).}
The only good CV candidate that was not confirmed spectroscopically is
Star~10 = PC1-V36. This object was highlighted as a possible CV in
Paper~I, based on 
its FUV brightness, variability and possible match to an X-ray
<source.\footnote{The status of Star~10=PC1-V36 as an X-ray
emitter remains uncertain: see the ``note added'' in Paper~I, as well
as Edmonds et al. (2003ab) and Heinke et al. (2005).}
As discussed in Section~\ref{sec:star10}, we now find that Star~10 is
an even more exotic system. 

Second, we have detected the secondary stars in two of our CVs. In
Star~1~=~AKO~9, the donor is a subgiant, whereas in Star~20~=~V2, it
is a lower-MS star. Moreover, our SED fit to Star~3~=~V1 constrains
the secondary in this systems to be a lower-MS star also. In principle,
the distribution of donor properties in the CV population is a 
strong constraint on theoretical models. For example, di Stefano \&
Rappaport (1994) predict quite a sizeable population of
bright, long-period CVs with evolved donors (like AKO~9) in 47~Tuc,
whereas such systems are relatively rare in the simulations of
Ivanova et al. (2006). As the number of known GC CVs grows, the
distributions of orbital period and donor spectral type will become
a powerful way to test models of CV formation and evolution.

Third, we estimated in Paper~I that $\simeq 15$~CVs should be
contained in our FUV imaging survey, based on the predictions by di
Stefano \& Rappaport (1994). Thus theory and observation would agree
perfectly if essentially all gap objects in Figure~\ref{fig:cmd} were
CVs. Even though our  spectroscopic sample comprises only half (8/16)
of the photometrically identified gap objects, it is clear that this
scenario is too optimistic. More specifically, even if we include the
marginal detection of line emission in Star~17, only half (4/8) of the
gap objects in the spectroscopic sample are confirmed as CVs. Is it 
possible that some or all of the other four gap objects in the
spectroscopic sample are nevertheless CVs that we have caught in a
state of weak or absent line emission? We believe this is extremely
unlikely. The spectroscopic/SED classifications for three of these gap
objects 
convincingly show that they are {\em not} CVs: Star~4 is a He~WD,
Star~10 is a different kind of exotic binary containing a stripped
subgiant; Star~27 is an SMC interloper. This leaves Star~15 as the
only other viable candidate. However, the possibility that it is
really a CV seems remote, given the lack of X-ray emission and FUV
variability (Paper~I), as well as the possibility that the match
between FUV and optical sources could be a chance coincidence
(Section~\ref{sec:star15}). 

Overall then, our results suggest that CVs comprise no more than 
$\sim 50$\% of the population in the gap, at least among the brighter objects. 
Nevertheless, taken at face value, this implies that predicted and observed 
numbers differ by ``only'' a factor of about two or three. However, this 
conclusion is not yet completely robust, both due to the small number
statistics involved, and because different theoretical models predict
different CV luminosity functions and hence different numbers of
observable systems (di Stefano \& Rappaport 1994; Ivanova et al. 2006).

\subsection{The More Exotic, the Better...}
\label{sec:exotica}

We finally want to briefly highlight what are arguably the three most
exotic objects we have uncovered. These are Star~2 (a BS-WD binary),
Star~4 (a He WD) and Star~10~=~PC1-V36 (a binary 
containing a dark primary and stripped subgiant secondary). We have
already outlined possible formation scenarios for these systems in
Sections~\ref{sec:star2}, \ref{sec:star4}, and \ref{sec:star10}, so we
will not repeat these here. Instead, we just note that our FUV survey
has turned out to be an efficient way of finding and 
classifying such remarkable systems. The gap 
between the WD and main sequences in the FUV-optical CMD seems to be a
particularly fertile hunting ground in this respect: both Stars~4
and 10 are found amongst only 8 gap sources in our spectroscopic
sample.

Aside from being intrinsically interesting, the most extreme objects
in GCs are important precisely because they are probably quite hard to
form. Star~10 may be a good example of this. The very existence of
such a system may point to a single, specific formation
mechanism. This contrasts with the 
relatively more common types of stellar exotica, for which there is
usually more than one formation channel (for an example in the CV
context, see Figure~7 in Ivanova et al. 2006). Thus a single unique
object can be the smoking gun that establishes the importance of a
particular dynamical process in GCs. 

\section{Summary \& Conclusions}

We have presented and analysed FUV spectroscopy for 48 blue objects in
the core of 47~Tucanae. For the 12 most interesting and representative
cases, we have also assembled broad-band, FUV-NIR spectral energy
distributions in order to determine their nature. Our main results are
as follows:

\begin{enumerate}

\item We have spectroscopically confirmed three previously known or
  suspected CVs in the cluster core (V1, V2 and AKO~9). For two of
  these (V2 and AKO~9), we have also found photometric evidence of
  dwarf nova eruptions in addition to those that were previously
  known. 

\item Only one other source -- Star~17, a ``gap object'' located
  between the WD and main sequences in the CMD -- exhibits marginal evidence for line
  emission in its spectrum. Thus the gap region is not exclusively
  populated by CVs, nor are these systems common amongst objects near
  the top of the WD cooling sequence. Nevertheless, predicted and
  observed numbers of CVs agree to within a factor of about 2-3. 

\item We have discovered a hot ($T_{eff} \simeq 8700$~K), large ($R
  \simeq 0.6~R_{\odot}$), but very low-mass ($M \simeq 0.06~
  M_{\odot}$) secondary star in a previously known 0.8~d binary
  system. This exotic object, Star~10, fills or nearly fills its
  Roche-lobe and completely dominates the binary's FUV-NIR output. We suggest that
  this star is the remains of a subgiant that has been almost
  completely stripped of its envelope. This stripping could have
  occurred as a result of mass
  transfer in the binary system, or during the dynamical interaction that
  actually formed the system. Since the stripped secondary must
  be in a short-lived evolutionary state, this object may represent
  the  ``smoking gun'' of a recent dynamical encounter. 

\item We have also found a Helium WD in 47~Tuc (Star~4). This is only
  the second optically detected such object in this cluster, and the first
  outside an MSP system. The He WD could have been formed in a
  primordial binary, or in a binary system formed in an exchange
  encounter involving a massive NS or WD. No sign of a companion is seen
  in our data.

\item We have discovered a bright BS with a young WD companion
  (Star~2). This is the only BS-WD binary known in any GC. However,
  the apparent rarity of such objects might be a selection
  effect. Since BS are much brighter than WDs at optical wavelengths, such binaries have been
  hard to find until now. In our FUV spectrum, the existence of a hot
  WD companion to the BS can be inferred from a strong FUV excess at the
  shortest wavelengths. 

\item In addition to the objects already discussed, we have found two
  more candidate WD binary systems. In one of these (Star~15), the
  apparent companion is a MS star, but this composite object may well
  be the product of a chance coincidence. However, the other system
  (Star~7) {\em is} likely to be a genuine binary, and the companion
  in this case is a sub-giant.

\item We have used the number of WD binary systems we have found to
  place a crude constraint on the WD binary fraction in the core of
  47~Tuc. We find $f_{bin} = 15\%~^{+17\%}_{-9\%}~{\rm stat}~^{+8\%}_{-7\%}~{\rm
    sys}$. Much stronger constraints can in principle be obtained from
  larger WD samples constructed from FUV surveys.

\item An SED fit to the optically brightest BS in our spectroscopic
  sample suggests that the mass of this star may exceed twice the
  turn-off mass. However, the uncertainties on the mass estimate are
  too large for this to be conclusive. There is still no definitive
  example of such a ``supermassive'' BS in any GC. 

\end{enumerate}

Overall, we feel that the present study is an excellent illustration
of the wide range of stellar exotica that are lurking in the cores of
GCs. In particular, the gap region of the CMD seems to harbour quite a 
variety of weird and wonderful objects. Most of these systems will
have undergone significant dynamical encounters or perhaps have even
been formed in one. All of this reenforces the motivation for studying
these systems. 

We also hope to have shown that FUV surveys -- and particularly
multi-object slitless FUV specroscopy -- provide an excellent and 
efficient way of achieving this goal. The power of this approach is
greatest, however, when it is used in combination with data spanning
the widest possible range in wavelength (e.g. the FUV-NIR range). With
such panchromatic data sets, we can determine the nature of most of
the exotic stellar systems that are hiding in GC cores.

\acknowledgments
We would like to thank the anonymous referee of this paper for a very
insightful and helpful report. 
Based on observations made with the NASA/ESA Hubble Space Telescope,
obtained at the Space Telescope Science Institute, which is operated
by the Association of Universities for Research in Astronomy, Inc., 
under NASA contract NAS 5-26555. All FUV observations are associated 
with program \#8279. Support for program \#8279 was provided by NASA
through a grant from the Space Telescope Science Institute, which is
operated by the Association of Universities for Research in Astronomy,
Inc., under NASA contract NAS 5-26555. C.K. and A.D. acknowledge
support from STFC through rolling grant
PP/D001013/1. J.M.A. acknowledges support from  the Spanish Government 
through grants AYA2004-08260-C03, 
AYA2004-05395, and AYA2007-64052 and from FEDER funds".

%% To help institutions obtain information on the effectiveness of their
%% telescopes, the AAS Journals has created a group of keywords for telescope
%% facilities. A common set of keywords will make these types of searches
%% significantly easier and more accurate. In addition, they will also be
%% useful in linking papers together which utilize the same telescopes
%% within the framework of the National Virtual Observatory.
%% See the AASTeX Web site at http://www.journals.uchicago.edu/AAS/AASTeX
%% for information on obtaining the facility keywords.

%% After the acknowledgments section, use the following syntax and the
%% \facility{} macro to list the keywords of facilities used in the research
%% for the paper.  Each keyword will be checked against the master list during
%% copy editing.  Individual instruments or configurations can be provided 
%% in parentheses, after the keyword, but they will not be verified.

{\it Facilities:} \facility{HST (STIS)}, \facility{HST (ACS)}

%\LongTables
%\begin{landscape}
%\tablewidth{1.0\textwidth}
%\begin{deluxetable*}{rllllllllll}
\begin{deluxetable}{rllllllllll}
\tablewidth{1.0\textwidth}
\tabletypesize{\tiny}
%\rotate
\tablecaption{Basic data on all FUV sources with useful spectra. Objects listed in bold at the top of the table are analyzed in detail in Section~\protect\ref{sec:sed}.\label{tab:specdat}}
\tablehead{
\colhead{ID} & 
\colhead{RA} & 
\colhead{Dec} & 
\colhead{$m_{FUV}$} & 
\colhead{$m_{F336W}$} &
\colhead{CMD} & 
\colhead{Sep} & 
\colhead{Var?} &
\colhead{PM?} & 
\colhead{Other Names} & 
\colhead{Spec/SED Class}\\
\colhead{(1)} &
\colhead{(2)} &
\colhead{(3)} &
\colhead{(4)} &
\colhead{(5)} &
\colhead{(6)} &
\colhead{(7)} &
\colhead{(8)} &
\colhead{(9)} &
\colhead{(10)} &
\colhead{(11)} 
}
\startdata
\bf{1}    & \bf{00:24:04.915}  & \bf{-72:04:55.40}  & \bf{15.62}  & \bf{17.99} & \bf{Gap}  & \bf{0.95}    & \bf{Yes}	& \bf{\ldots} & \bf{AKO 9, W36, PC1-V11,M055581}    & \bf{CV (WD-SG)}                               \\
\bf{2}    & \bf{00:24:06.541}  & \bf{-72:05:03.20}  & \bf{17.37}  & \bf{15.72} & \bf{BS}   & \bf{0.74}    & \bf{Yes}	& \bf{\ldots} & \bf{M048212}	                    & \bf{Binary (BS-WD)}                           \\       
\bf{3}    & \bf{00:24:04.248}  & \bf{-72:04:57.99}  & \bf{17.53}  & \bf{19.11} & \bf{Gap}  & \bf{1.05}    & \bf{Yes}	& \bf{\ldots} & \bf{V1, X9, W42, PC1-V47, M053129}  & \bf{CV (WD-MS)}                               \\                  
\bf{4}    & \bf{00:24:06.353}  & \bf{-72:04:49.17}  & \bf{17.70}  & \bf{19.96} & \bf{Gap}  & \bf{1.00}    & \bf{No} 	& \bf{\ldots} & \bf{M061563}	                    & \bf{He WD}                                    \\       
\bf{5}    & \bf{00:24:05.557}  & \bf{-72:05:02.15}  & \bf{18.03}  & \bf{20.92} & \bf{WD}   & \bf{1.01}    & \bf{No} 	& \bf{\ldots} & \bf{M049123}	                    & \bf{WD}                                       \\  
\bf{7}    & \bf{00:24:05.794}  & \bf{-72:04:57.13}  & \bf{18.26}  & \bf{20.85} & \bf{WD}   & \bf{1.07}    & \bf{No} 	& \bf{\ldots} & \bf{\ldots}	                    & \bf{Binary (WD-MS)}                           \\
\bf{10}   & \bf{00:24:06.489}  & \bf{-72:04:52.27}  & \bf{18.35}  & \bf{18.07} & \bf{Gap}  & \bf{0.59}    & \bf{No} 	& \bf{Yes}    & \bf{W75?, PC1-V36, DM-3, M058539}   & \bf{Binary (??-Stripped SG)}                  \\
\bf{15}   & \bf{00:24:06.563}  & \bf{-72:05:04.38}  & \bf{18.63}  & \bf{18.03} & \bf{Gap}  & \bf{1.29}    & \bf{No} 	& \bf{\ldots} & \bf{M047149}	                    & \bf{Binary? (WD-SG)\tablenotemark{c}}                           \\  
\bf{17}   & \bf{00:24:06.390}  & \bf{-72:05:07.37}  & \bf{18.77}  & \bf{18.61} & \bf{Gap}  & \bf{1.20}    & \bf{No} 	& \bf{\ldots} & \bf{M044437}	                    & \bf{Binary? (WD-SG)\tablenotemark{c}; CV?}                      \\
\bf{20}   & \bf{00:24:05.992}  & \bf{-72:04:56.11}  & \bf{18.99}  & \bf{19.94} & \bf{Gap}  & \bf{1.07}    & \bf{Yes}	& \bf{\ldots} & \bf{V2, X19, W30, PV1-V53, M054898} & \bf{CV (WD-MS)}                               \\ 
\bf{27}   & \bf{00:24:06.850}  & \bf{-72:04:53.35}  & \bf{19.41}  & \bf{20.66} & \bf{Gap}  & \bf{0.75}    & \bf{No}     & \bf{\ldots} & \bf{M057483}	                    & \bf{SMC Interloper}                           \\       
\bf{999}  & \bf{00:24:04.009}  & \bf{-72:04:51.24}  & \bf{17.61}  & \bf{15.72} & \bf{BS}   & \bf{\ldots}  & \bf{\ldots} & \bf{\ldots} & \bf{M059651\tablenotemark{a}}       & \bf{BS}                                       \\ 
     6    &     00:24:08.459   &     -72:04:50.58   &     18.15   &     \ldots &     NoOpt &     \ldots   &     No 	& \ldots      & \ldots                              & \ldots                                        \\
     8    &     00:24:04.912   &     -72:04:51.67   &     18.27   &     \ldots &     NoOpt &     \ldots   &     No 	& \ldots      & \ldots                              & \ldots	                                    \\
     9    &     00:24:07.443   &     -72:05:07.43   &     18.33   &     21.10  &     WD    &     1.43     &     No 	& \ldots      & \ldots                              & \ldots	                                    \\
     12   &     00:24:07.919   &     -72:04:52.99   &     18.41   &     15.87  &     BS    &     0.48     &     Yes	& Yes         & M057848	                            & \ldots	                                    \\       
     13   &     00:24:04.890   &     -72:05:03.61   &     18.46   &     21.36  &     WD    &     1.24     &     No 	& \ldots      & \ldots                              & \ldots	                                    \\
     14   &     00:24:05.554   &     -72:04:53.94   &     18.51   &     16.15  &     BS    &     0.44     &     No 	& Yes         & DM-9, M056964                       & \ldots	                                    \\        
     16   &     00:24:06.511   &     -72:04:50.28   &     18.76   &     16.34  &     BS    &     0.43     &     Yes	& Yes         & M060466\tablenotemark{b}            & \ldots	                                    \\  
     18   &     00:24:05.266   &     -72:04:59.02   &     18.99   &     21.74  &     WD    &     1.00     &     No 	& \ldots      & \ldots                              & \ldots	                                    \\
     23   &     00:24:06.398   &     -72:04:53.92   &     19.10   &     16.26  &     BS    &     1.34     &     No 	& Yes         & DM-5,M056966\tablenotemark{b}       & \ldots	                                    \\       
     24   &     00:24:06.122   &     -72:04:51.84   &     19.14   &     \ldots &     NoOpt &     \ldots   &     No 	& \ldots      & \ldots                              & \ldots	                                    \\      
     25   &     00:24:08.023   &     -72:05:06.30   &     19.18   &     \ldots &     NoOpt &     \ldots   &     No 	& \ldots      & \ldots                              & \ldots	                                    \\      
     26   &     00:24:07.456   &     -72:04:57.08   &     19.25   &     \ldots &     NoOpt &     \ldots   &     No 	& \ldots      & \ldots                              & \ldots	                                    \\      
     29   &     00:24:07.933   &     -72:04:57.06   &     19.48   &     22.14  &     WD    &     1.18     &     No 	& \ldots      & M053994                             & \ldots	                                    \\  
     30   &     00:24:07.024   &     -72:04:49.13   &     19.51   &     21.93  &     WD    &     0.63     &     No 	& \ldots      & \ldots	                            & \ldots	                                    \\       
     31   &     00:24:04.246   &     -72:04:51.39   &     19.58   &     22.29  &     WD    &     0.75     &     No 	& \ldots      & \ldots                              & \ldots	                                    \\      
     32   &     00:24:03.646   &     -72:05:04.15   &     19.59   &     \ldots &     NoOpt &     \ldots   &     No 	& \ldots      & \ldots                              & \ldots	                                    \\      
     33   &     00:24:05.986   &     -72:04:58.37   &     19.62   &     \ldots &     NoOpt &     \ldots   &     No 	& \ldots      & \ldots                              & \ldots	                                    \\      
     36   &     00:24:04.171   &     -72:05:02.45   &     19.71   &     \ldots &     NoOpt &     \ldots   &     No 	& \ldots      & \ldots                              & \ldots	                                    \\      
     37   &     00:24:05.234   &     -72:05:02.06   &     19.72   &     \ldots &     NoOpt &     \ldots   &     No 	& \ldots      & \ldots                              & \ldots	                                    \\      
     38   &     00:24:07.375   &     -72:04:51.13   &     19.77   &     22.33  &     WD    &     0.82     &     No 	& \ldots      & \ldots		                    & \ldots	                                    \\       
     42   &     00:24:05.982   &     -72:04:55.34   &     19.99   &     \ldots &     NoOpt &     \ldots   &     No 	& \ldots      & \ldots                              & \ldots	                                    \\      
     46   &     00:24:07.228   &     -72:05:01.21   &     20.16   &     \ldots &     NoOpt &     \ldots   &     No 	& \ldots      & \ldots                              & \ldots	                                    \\      
     47   &     00:24:04.623   &     -72:04:47.95   &     20.20   &     \ldots &     NoOpt &     \ldots   &     No 	& \ldots      & \ldots                              & \ldots	                                    \\      
     49   &     00:24:06.085   &     -72:05:06.72   &     20.27   &     \ldots &     NoOpt &     \ldots   &     No 	& \ldots      & \ldots                              & \ldots	                                    \\      
     52   &     00:24:05.964   &     -72:04:48.94   &     20.40   &     16.73  &     BS    &     0.52     &     No 	& Yes         & M061787\tablenotemark{a}            & \ldots	                                    \\  
     54   &     00:24:05.736   &     -72:04:53.31   &     20.47   &     16.80  &     BS    &     0.27     &     No 	& Yes         & M057553                             & \ldots	                                    \\  
     59   &     00:24:07.383   &     -72:05:05.76   &     20.55   &     \ldots &     NoOpt &     \ldots   &     No 	& \ldots      & \ldots                              & \ldots	                                    \\      
     66   &     00:24:04.299   &     -72:05:05.01   &     20.68   &     \ldots &     NoOpt &     \ldots   &     No 	& \ldots      & \ldots                              & \ldots	                                    \\      
     67   &     00:24:04.096   &     -72:04:51.27   &     20.69   &     \ldots &     NoOpt &     \ldots   &     No 	& \ldots      & \ldots                              & \ldots	                                    \\      
     73   &     00:24:08.222   &     -72:04:52.19   &     20.86   &     \ldots &     NoOpt &     \ldots   &     No 	& \ldots      & \ldots                              & \ldots	                                    \\      
     78   &     00:24:04.374   &     -72:04:58.26   &     20.99   &     \ldots &     NoOpt &     \ldots   &     No 	& \ldots      & \ldots                              & \ldots	                                    \\      
     79   &     00:24:08.240   &     -72:04:53.58   &     21.00   &     \ldots &     NoOpt &     \ldots   &     No 	& \ldots      & \ldots                              & \ldots	                                    \\      
     81   &     00:24:06.308   &     -72:05:05.27   &     21.05   &     \ldots &     NoOpt &     \ldots   &     No 	& \ldots      & \ldots                              & \ldots	                                    \\      
     82   &     00:24:06.800   &     -72:05:00.12   &     21.06   &     \ldots &     NoOpt &     \ldots   &     No 	& \ldots      & \ldots                              & \ldots	                                    \\      
     84   &     00:24:07.678   &     -72:04:55.88   &     21.09   &     \ldots &     NoOpt &     \ldots   &     No 	& \ldots      & \ldots                              & \ldots	                                    \\      
     85   &     00:24:06.341   &     -72:04:51.16   &     21.17   &     \ldots &     NoOpt &     \ldots   &     No 	& \ldots      & \ldots                              & \ldots                                        
\enddata
\tablenotetext{a}{The master image of McLaughlin et al. (2006) is saturated and/or confused in the vicinity of this object.}
\tablenotetext{b}{This object has a near neighbour that was not detected in McLaughlin et al. (2006).}
\tablenotetext{c}{The binary status of this system is not certain due to the possibility of a chance superposition.}
%\tabletypesize{\scriptsize}
%\tabletypesize{\tiny}
%% Text for table notes should follow after the \enddata but before
%% the \end{deluxetable}. Make sure there is at least one \tablenotemark
%% in the table for each \tablenotetext.
\tablecomments{Column definitions:\\
(1) Spectrosopic identification number; c.f. Figure~\protect\ref{fig:images}\\
(2) Right ascension (J2000); boresight corrected to match positions in Heinke et al. (2005).\\
(3) Declination (J2000); boresight corrected to match positions in Heinke et al. (2005).\\
(4) STIS/FUV-MAMA/F25QTZ magnitude (STMAG system).\\
(5) WFPC2/PC/F336W magnitude (STMAG system).\\
(6) CMD-based classification; see Section~\protect\ref{sec:cmd} and Figure~\protect\ref{fig:cmd}.\\ 
(7) Offset between FUV and F336W position in STIS/FUV-MAMA pixels (1 pix $\simeq$\ 0.025\arcsec).\\
(8) Variable in FUV photometry? (See Paper~I for details.)\\
(9) Proper motion consistent with cluster membership? (Only if available in McLaughlin et al. [2006].)\\
(10) Alternative names are taken from the following catalogues:\\
\hspace*{0.5cm} AKO~n = Star n in Auri\`ere, Koch-Miramond \& Ortolani (1989); \\
\hspace*{0.5cm} Vn = Variable star n as defined by Paresce, De Marchi \& Ferraro (1992) and Paresce \& De Marchi(1994)\\
\hspace*{0.5cm} Wn = Star n in Grindlay et al. (2001a) and Heinke et al. (2005); \\
\hspace*{0.5cm} PC1-Vn = Variable star n on the PC1 chip in Albrow et al. (2001);\\
\hspace*{0.5cm} Mn = Star n in the master catalogue of McLaughlin et al. (2006);\\
\hspace*{0.5cm} DM-n = Star 104-n in De Marco et al. (2005).\\
(11) Final classification based on FUV spectrum and FUV-NIR SED (Section~\protect\ref{sec:sed}). Notation as in text, except G=subgiant. 
}
%\vspace*{0.5cm}
%
%\tablenotetext{a}{Spectrosopic identification number; c.f. Figure~\protect\ref{fig:images}.}
%\tablenotetext{b}{Right ascension (J2000); boresight corrected to match positions in Heinke et al. (2005).}
%\tablenotetext{c}{Declination (J2000); boresight corrected to match positions in Heinke et al. (2005).}
%\tablenotetext{d}{STIS/FUV-MAMA/F25QTZ magnitude (STMAG system).}
%\tablenotetext{e}{WFPC2/PC/F336W magnitude (STMAG system).}
%\tablenotetext{f}{CMD-based classification; see Section~\protect\ref{sec:cmd} and Figure~\protect\ref{fig:cmd}.}
%\tablenotetext{g}{Offset between FUV and F336W position in STIS/FUV-MAMA pixels (1 pix $\simeq$\ 0.025\arcsec).}
%\tablenotetext{h}{Variable in FUV photometry? (See Paper~I for details.)}
%\tablenotetext{i}{AKO~n = Star n in Auri\'ere, Koch-Miramond \& Ortolani (1989); \\
                  %Vn = Star n in Paresce, De Marchi \& Ferraro (1992); \\
                  %Wn = Star n in Grindlay et al. (2001) and Heinke et al. (2005); \\
                  %PC1-Vn = Star n on the PC1 chip in Albrow et al. (2001);\\
                  %Mn = Star n in the master catalogue of McLaughlin et al. (2006);\\
                  %DM-n = Star n in the 47 Tuc data of De Marco et
                  %al. (2005).}                                                                                             
%\tablenotetext{j}{Final classification based on the analysis in Section~\protect\ref{sec:sed}.}
%\end{deluxetable*}
%\end{landscape}
\end{deluxetable}

\end{document}